\documentclass[amsmath,preprintnumbers,11pt]{revtex4-1}
\usepackage{amsmath,amsfonts,bm}
\usepackage{graphicx}
\usepackage{color}

\usepackage{verbatim}

\begin{document}
\title{Escorted Free Energy Simulations}
\author{Suriyanarayanan Vaikuntanathan$^1$ and Christopher Jarzynski$^{1,2}$}
\affiliation{$^1$Chemical Physics Program, Institute for Physical Science and Technology,University of Maryland, College Park, MD 20742\\ 
$^2$Department of Chemistry and Biochemistry, University of Maryland, College Park, MD 20742}
\begin{abstract}
We describe a strategy to improve the efficiency of free energy estimates by reducing dissipation in nonequilibrium Monte Carlo simulations.
This strategy generalizes the targeted free energy perturbation approach [{\textit {Phys. Rev. E.}} {\bf 65}, 046122, 2002] to nonequilibrium switching simulations, and involves generating artificial, ``escorted'' trajectories by coupling the evolution of the system to updates in external work parameter. Our central results are: (1) a generalized fluctuation theorem for the escorted trajectories, and (2) estimators for the free energy difference $\Delta F$ in terms of these trajectories. We illustrate the method and its effectiveness on model systems.

\end{abstract}
\maketitle

\section{Introduction}

The computation of free energy differences is an essential component of computer studies of biological, chemical, and molecular processes, with applications to topics such as phase coexistence and phase equilibria, ligand binding events, and solvation of small molecules~\cite{Frenkel,Chipot2007}. Given the importance of free energy calculations in computational thermodynamics, there is a need for robust, efficient and accurate methods to estimate free energy differences. 

\begin{figure}[tbp]         
\includegraphics[scale=0.4,angle=0]{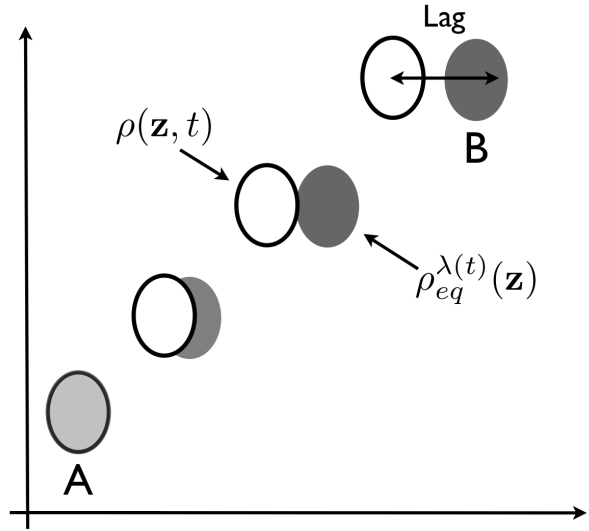} 
\caption{
The axes schematically represent configuration space (${\bf z}$-space).
The unshaded ovals denote the statistical state of the system, $\rho({\bf z},t)$, and the shaded ovals denote the equilibrium state, $\rho^{\lambda(t)}_{eq}({\bf z})$,  corresponding to the value of external parameter, $\lambda(t)$, at various instants of time.
As the work parameter $\lambda$ is switched from $A$ to $B$, a lag builds up as the state of the system, $\rho({\bf z},t)$, pursues the equilibrium distribution corresponding to the changing work parameter, $\rho^{\lambda(t)}_{eq}({\bf z})$. 
}
 \label{lag.sketch}
\end{figure}

In a standard formulation of the free energy estimation problem, we consider two equilibrium states of a system, corresponding to the same temperature $T$ but different values of an external parameter, $\lambda=A,B$, and we are interested in the free energy difference between the two states, $\Delta F= F_B-F_A$. While many widely used free energy estimation methods, such as thermodynamic integration and free energy perturbation rely on equilibrium sampling, there has been considerable interest in methods for estimating $\Delta F$ that make use of nonequilibrium simulations~\cite{Frenkel,Chipot2007}. In the most direct implementation of this approach, a number of independent simulations are performed in which the external parameter is varied at a finite rate from $\lambda=A$ to $\lambda=B$, with initial conditions sampled from the equilibrium state $A$. The free energy difference $\Delta F$ can then be estimated using the nonequilibrium work relation~\cite{CJ:Equality,CJ:MasterEquation}
\begin{equation}
\label{NEW}
e^{-\beta \Delta F} = \langle e^{-\beta W} \rangle
\end{equation}
where $W$ denotes the work performed on the system during a particular realization (i.e. simulation) of the process, angular brackets $\langle\dots\rangle$ denote an average over the realizations of the process and $\beta=1/T$. 
In principle, this approach allows one to compute $\Delta F$ from trajectories of arbitrarily short duration. However, the number of realizations required to obtain a reliable estimate of $\Delta F$ grows rapidly with the dissipation, $\langle W_{diss}\rangle \equiv \langle W\rangle -\Delta F$, that accompanies fast switching simulations~\cite{Kofke,CJ:Rareevents,Gore} . The dissipation is positive as a consequence of the second law of thermodynamics, and reflects the lag that builds up as the system pursues -- but is unable to keep pace with -- the equilibrium distribution corresponding to the continuously changing parameter $\lambda$~\cite{Lag1,Lag2,Hermans91,SVCJ_lag}.  This idea is illustrated schematically in Fig \ref{lag.sketch}.

In Ref.~\cite{escorted}, we described a strategy to improve the efficiency of free energy estimates obtained with nonequilibrium molecular-dynamics simulations. This strategy involved adding non-physical terms to the equations of motion, to reduce the lag and therefore the dissipation. As illustrated in Ref.~\cite{escorted} using a simple model system, when these terms successfully ``escorted'' the system through a near-equilibrium sequence of states, the convergence of the free energy estimate improved dramatically.  In the present paper we extend these results to simulations evolving according to Monte Carlo dynamics.
We then show that the escorted trajectories satisfy a fluctuation theorem, and we discuss and illustrate the application of this result to the estimation of free energy differences.

In Section \ref{Theory} we introduce escorted nonequilibrium switching simulations for systems evolving according to Monte Carlo dynamics. The approach we take here is motivated by previous work~\cite{escorted,CJ:Targeting,BijectiveTFEP,miller:7035} and involves generating artificial, or ``escorted'', trajectories, Eq. \ref{Map.scheme}, by modifying the dynamics with terms that directly couple the evolution of the system to changes in the external parameter. The central result of this section is an identity for $\Delta F$ in terms of these escorted trajectories, Eq. \ref{Map.discreteF}. In Section \ref{FT} we extend this result by showing that these trajectories satisfy a fluctuation relation analogous to Crooks's fluctuation relation{~\cite{Crooksfluctuation,Crooks1998,Crooks99}}. 
This in turn allows us to combine our approach with Bennett's acceptance ratio method~\cite{Bennett} which provides an optimal, asymptotically unbiased estimator, Eq. \ref{Bennett:FT}, for $\Delta F$~\cite{Bennett:Likelihood}. In Section \ref{Merit}, we show that while Eqs. \ref{Map.discreteF} and \ref{Bennett:FT} are identities for all escorted simulations, they are particularly effective as estimators of $\Delta F$ when the modified dynamics successfully reduce the lag described above. In particular, if these terms eliminate the lag entirely, then Eqs. \ref{Map.discreteF} and \ref{Bennett:FT} provide perfect (zero variance) estimators: W = $\Delta F$ for every realization. Finally in Section \ref{Examples}, we illustrate the effectiveness of our approach on two model systems.

\section{Escorted nonequilibrium simulations}
 \label{Theory}
Consider a system whose energy is given by a classical hamiltonian, $H_\lambda({\bf z})$, where ${\bf z}$ denotes a microstate, that is a point in the $D$-dimensional configuration space of the system,~\footnote{As is usually the case with Monte Carlo simulations, we do not include momenta in the microstate.}
and $\lambda$ is an external work parameter. At a temperature $\beta^{-1}$, the equilibrium state of this system is described by the distribution
\begin{equation}
\label{equilibrium}
\rho^{\lambda}_{eq}({\bf z})=\frac{e^{-\beta H_{\lambda}({\bf z})}}{Z_{\lambda}}
\end{equation}
with the free energy  $F_\lambda=-\beta^{-1}\ln Z_{\lambda}$. We wish to compute the free energy difference $\Delta F= F_B-F_A$ between two equilibrium states at the same temperature, $\beta^{-1}$, but different values of the work parameter, $\lambda=A,B$. 

To estimate the value of $\Delta F$, we assume we have at our disposal a discrete-time Monte Carlo algorithm, parametrized by the value of $\lambda$ and defined by the transition probability $P_\lambda({\bf z}\vert{\bf z}_0)$: if ${\bf z}_0$ represents the microstate of  the system at one time step, then the next microstate ${\bf z}$ is sampled randomly from $P_{\lambda}({\bf z}\vert {\bf z}_0)$.
We assume this algorithm satisfies the conditions of {\it detailed balance},
\begin{equation}
\label{D.Balance}
\frac{P_{\lambda}({\bf z} | {\bf z}_0)}{P_{\lambda}({\bf z}_0 | {\bf z})}=\frac{e^{-\beta H_{\lambda}({\bf z})}}{e^{-\beta H_{\lambda}({\bf z}_0)}} 
\end{equation}
and {\it ergodicity}~\cite{Kampen}.
Routinely used Monte Carlo schemes such as the Metropolis algorithm~\cite{Frenkel} satisfy these conditions.
Eq.~\ref{D.Balance} implies the somewhat weaker condition of {\it balance},
\begin{equation}
\label{balance}
\int d{\bf z}_0 \, P_{\lambda}({\bf z} | {\bf z}_0) \, 
e^{-\beta H_{\lambda}({\bf z}_0)}
=
e^{-\beta H_{\lambda}({\bf z})}
\end{equation}
which we will use in the analysis below.
With this Monte Carlo algorithm in place, we first describe a standard procedure for estimating $\Delta F$ using nonequilibrium simulations, Eqs.~\ref{scheme}-\ref{discreteNEW} below, and then we introduce our modified version of this approach.

Imagine a process in which the system is initially prepared in equilibrium, at $\lambda=A$ and temperature $\beta^{-1}$, and then the system evolves under the Monte Carlo dynamics described above, as the value of $\lambda$ is switched from $A$ to $B$ in $N$ steps according to some pre-determined protocol.
This evolution generates a trajectory ${\bf \gamma} = \{ {\bf z}_0,{\bf z}_1,\dots,{\bf z}_{N-1}\}$ that can be represented in more detail using the notation
\begin{equation}
\label {scheme}
	[{\bf z}_0,\lambda_0]\Rightarrow[{\bf z}_0,\lambda_1]\rightarrow[{\bf z}_1,\lambda_1]\Rightarrow\cdots
	\rightarrow[{\bf z}_{N-1},\lambda_{N-1}]
	\Rightarrow[{\bf z}_{N-1},\lambda_N] .
\end{equation}
Here, the symbol $\Rightarrow$ denotes an update in the value of $\lambda$, with the microstate held fixed, while $\rightarrow$ denotes a Monte Carlo step at fixed $\lambda$,
e.g.\ the microstate ${\bf z}_1$ is sampled from the distribution $P_{\lambda_1}({\bf z}_1 \vert {\bf z}_0)$.
Moreover,
\begin{equation}
\label{eq:terminalValues}
\lambda_0\equiv A
\qquad,\qquad
\lambda_N \equiv B ,
\end{equation}
and the initial point ${\bf z}_0$ is sampled from $\rho^A_{eq}({\bf z}_0)$.

Because it is specified by the sequence of microstates ${\bf z}_0,\cdots {\bf z}_{N-1}$, the trajectory $\gamma$ can be viewed as a point in a $DN$-dimensional trajectory space, with $d\gamma = d{\bf z}_0 \cdots d{\bf z}_{N-1}$.
For the process described in the previous paragraph, the probability density for generating this trajectory is
\begin{equation}
\label{gammaP}
p[\gamma]=
P_{\lambda_{N-1}}({\bf z}_{N-1} \vert {\bf z}_{N-2}) \cdots
P_{\lambda_2}({\bf z}_2 \vert {\bf z}_1) \, P_{\lambda_1}({\bf z}_1 \vert {\bf z}_0) \, 
\rho^{A}_{eq}({\bf z}_0)
\end{equation}
where the factors $P_{\lambda_i}({\bf z}_i \vert {\bf z}_{i-1})$ in this equation (read from right to left) correspond to the symbols $\rightarrow$ in Eq.~\ref{scheme} (read from left to right).
The work performed on the system during this process is the sum of energy changes due to updates in $\lambda$,~\cite{Reinhardt1992,Hunter1993,CJ:MasterEquation,Crooks1998}
\begin{equation}
\label{NEWwork}
W[\gamma]= \sum_{i=0}^{i=N-1} \delta W_i\equiv \sum_{i=0}^{i=N-1} \left[ H_{\lambda_{i+1}}({\bf z}_i)-H_{\lambda_{i}}({\bf z}_i) \right] .
\end{equation}
Using Eqs. \ref{D.Balance},  \ref{gammaP} and \ref{NEWwork}, we arrive at the nonequilibrium work relation for Monte Carlo dynamics~\cite{CJ:MasterEquation,Crooks1998}
\begin{equation}
\label{discreteNEW}
\langle e^{-\beta W} \rangle
\equiv \int d\gamma \, p[\gamma] e^{-\beta W[\gamma]}
=e^{-\beta \Delta F}.
\end{equation}
Thus we can estimate $\Delta F$ by repeatedly performing simulations to generate trajectories of the sort described by Eq.~\ref{scheme}, computing the work associated with each trajectory, Eq.~\ref{NEWwork}, and finally constructing the exponential average, Eq.~\ref{discreteNEW}.
As mentioned in the Introduction, however, this average converges poorly when the process is highly dissipative.

To address the issue of poor convergence, let us now assume that for every integer $0 \le i < N$, we have a deterministic function $M_i:{\bf z}\rightarrow{\bf z^{\prime}}$ that takes any point ${\bf z}$ in configuration space and maps it to a point ${\bf z}^\prime$.
We assume that each of these functions is invertible ($M_i^{-1}$ exists), but otherwise the functions are arbitrary.
These $M_i$'s then constitute a set of {\it bijective mappings}, which we use to modify the procedure for generating trajectories, as follows.
When the value of the work parameter is switched from $\lambda_i$ to $\lambda_{i+1}$, the configuration space coordinates are simultaneously subjected to the mapping $M_i$. Eq. \ref{scheme} then becomes
\begin{equation}
\label {Map.scheme}
	[{\bf z}_0,\lambda_0]\stackrel{M_0}{\Rightarrow}[{\bf z}_0^{\prime},\lambda_1]\rightarrow[{\bf z}_1,\lambda_1] 
	\stackrel{M_1}{\Rightarrow} \cdots
	\rightarrow [{\bf z}_{N-1},\lambda_{N-1}]
	\stackrel{M_{N-1}}{\Rightarrow}[{\bf z}_{N-1}^\prime,\lambda_N]
\end{equation}
where
\begin{equation}
\label{mapping}
{\bf z}_i^\prime \equiv M_i({\bf z}_i) ,
\end{equation}
as indicated by the notation $\stackrel{M_i}{\Rightarrow}$.
(As before, the symbol $\rightarrow$ denotes a Monte Carlo move at fixed $\lambda$.)
The bijective maps effectively escort the system by directly coupling increments in $\lambda$ to changes in the microstate.
This is similar to the ``metric scaling'' approach introduced by Miller and Reinhardt~\cite{miller:7035}, in which each update in $\lambda$ is accompanied by a linear scaling of coordinates;
however, in the present paper we do not assume the $M_i$'s are linear in ${\bf z}$.

In the escorted trajectory (Eq. ~\ref{Map.scheme}), the system visits a sequence of $2N$ points in configuration space: the $N$ ``primary'' microstates ${\bf z}_0, \cdots {\bf z}_{N-1}$, alternating with the $N$ ``secondary'' microstates ${\bf z}_0^\prime, \cdots {\bf z}_{N-1}^\prime$.
Since each ${\bf z}_i^\prime$ is uniquely determined from ${\bf z}_i$ (Eq. ~\ref{mapping}), the sequence of primary microstates $\gamma = \{ {\bf z}_0, \cdots {\bf z}_{N-1} \}$ fully specifies the trajectory;
that is, trajectory space remains $DN$-dimensional, with $d\gamma = d{\bf z}_0 \cdots d{\bf z}_{N-1}$.
The probability density for generating a trajectory $\gamma$ is given by the following modification of Eq. ~\ref{gammaP}:
\begin{equation}
\label{Map.gammaP}
p[\gamma]=
P_{\lambda_{N-1}}({\bf z}_{N-1} \vert  {\bf z}_{N-2}^\prime)
\cdots 
P_{\lambda_2}({\bf z}_2 \vert  {\bf z}_1^\prime) \,
P_{\lambda_1}({\bf z}_1 \vert  {\bf z}_0^\prime) \,
\rho^{A}_{eq}({\bf z}_0) 
\end{equation}
Taking a cue from Refs~\cite{CJ:Targeting,miller:7035}, let us now define 
\begin{equation}
\label{Map.work}
W^\prime[\gamma]= \sum_{i=0}^{N-1} \delta W_i^\prime 
\equiv \sum_{i=0}^{N-1} \left[ H_{\lambda_{i+1}}({\bf z}_i^\prime)-H_{\lambda_{i}}({\bf z}_i)-\beta^{-1}\ln J_i({\bf z}_i) \right]
\end{equation}
where $J_i({\bf z}) = \vert \partial{\bf z}^\prime/\partial{\bf z}\vert$ is the Jacobian associated with the map $M_i:{\bf z}\rightarrow{\bf z}^\prime$.
Averaging $\exp(-\beta W^\prime[\gamma])$ over the ensemble of trajectories, we have
\begin{eqnarray} 
\label{Map.averageexpW}
\langle e^{-\beta W^\prime} \rangle &=&
\int d\gamma \,
p[\gamma] \nonumber \, e^{-\beta W^\prime[\gamma]} \\
=
\frac{1}{Z_{\lambda_0}}
\int d{\bf z}_{N-1}
&\cdots&
\int d{\bf z}_{0} \, e^{-\beta \sum_{i=0}^{N-1} \delta W_i^\prime}
 \,
 P_{\lambda_{N-1}}({\bf z}_{N-1} \vert  {\bf z}_{N-2}^\prime) 
 \dots 
 P_{\lambda_1}({\bf z}_1 \vert  {\bf z}_0^\prime) \,
 e^{-\beta H_{\lambda_0}({\bf z}_0)}
\end{eqnarray}
To evaluate this expression, we first identify all factors in the integrand that do not depend on ${\bf z}_0$ or ${\bf z}_0^\prime$, and we pull these outside the innermost integral, $\int d {\bf z}_0$, which gives us (for that integral):
\begin{eqnarray}
&~&\int \,d{\bf z}_0 \,
e^{-\beta\delta W_0^\prime}\,
P_{\lambda_1}({\bf z}_1 \vert {\bf z}_0^\prime) \,
e^{-\beta H_{\lambda_0}({\bf z}_0)} \\
&=&\int \,d{\bf z}_0\, J_0({\bf z}_0) \, 
P_{\lambda_1}({\bf z}_1|{\bf z}_0^\prime) \, e^{-\beta H_{\lambda_1}({\bf z}_0^\prime)}  \\
&=& \int \,d{\bf z}_0^\prime \,  P_{\lambda_1}({\bf z}_1|{\bf z}_0^\prime) \,  e^{-\beta H_{\lambda_1}({\bf z}_0^\prime)} 
 =  e^{-\beta H_{\lambda_1}({\bf z}_1)}
\end{eqnarray}
We have used Eq. ~\ref{Map.work} to get to the second line, followed by a change in the variables of integration to get to the third line, $d{\bf z}_0\, J_0({\bf z}_0) \rightarrow d{\bf z}_0^\prime$, and we have invoked Eq. ~\ref{balance} to arrive at the final result.
This process can be repeated for the integrals $\int d{\bf z}_1$ to $\int d{\bf z}_{N-2}$, which brings us to:
\begin{eqnarray}
\langle e^{-\beta W^\prime}\rangle &=&
\frac{1}{Z_{\lambda_0}}\int \,d{\bf z}_{N-1} \,
e^{-\beta \delta W_{N-1}^\prime} \,
e^{-\beta H_{\lambda_{N-1}}({\bf z}_{N-1})} \nonumber \\
&=&
\frac{1}{Z_{\lambda_0}}\int \,d{\bf z}_{N-1} \,
J_{N-1}({\bf z}_{N-1}) \, e^{-\beta H_{\lambda_N}({\bf z}_{N-1}^\prime)} \\
&=&
\frac{1}{Z_{\lambda_0}}\int \,d{\bf z}_{N-1}^\prime \,
e^{-\beta H_{\lambda_N}({\bf z}_{N-1}^\prime)}
= \frac{Z_{\lambda_N}}{Z_{\lambda_0}},
\nonumber
\end{eqnarray}
and therefore
\begin{equation}
\label{Map.discreteF}
\langle e^{-\beta W^\prime}\rangle = e^{-\beta\Delta F}
\end{equation}

Eq. \ref{Map.discreteF} is an identity for $\Delta F$ in terms of escorted trajectories, generated as per Eq.~\ref{Map.scheme}.
For the special case in which each mapping is the identity, $M_i=I$, we recover the usual scheme, Eq.~\ref{scheme}, and then Eq.~\ref{Map.discreteF} reduces to the nonequilibrium work relation, Eq.~\ref{discreteNEW}.
Following Miller and Reinhardt~\cite{miller:7035}, we will find it convenient to interpret $W^\prime$ as the work done during the switching process and simply denote it by $W$.  
As we will discuss in Section~\ref{Merit} below, when the mappings $\{ M_i \}$ are chosen so as to reduce the dynamic lag illustrated in Fig.~\ref{lag.sketch}, then the efficiency of the estimate of $\Delta F$ improves, often dramatically.

\section{Fluctuation Theorem}
\label{FT}

Let us now consider not only the switching process described by Eq.~\ref{Map.scheme}, which we will henceforth designate the {\it forward} process, but also its time-reversed analogue, the {\it reverse} process.
In the reverse process, the system is prepared in equilibrium at $\lambda=B$ and temperature $\beta^{-1}$. 
The work parameter is then switched to $\lambda=A$ in $N$ steps, following a sequence $\{ \tilde \lambda_{0}, \tilde \lambda_{1},\cdots,\tilde \lambda_N\}$ that is the reversal of the protocol used during the forward process:
\begin{equation}
\tilde \lambda_i\equiv \lambda_{N-i} 
\end{equation}
During the reverse process, changes in $\lambda$ are coupled to the system's evolution through the inverse mapping functions, $\tilde M_i \equiv M_{N-1-i}^{-1}$, generating a trajectory
\begin{equation}
\label {Map.scheme.R}
[\tilde{\bf z}_{N-1}^\prime,\tilde\lambda_N] \stackrel{\tilde M_{N-1}}{\Leftarrow} [\tilde{\bf z}_{N-1},\tilde\lambda_{N-1}] \leftarrow \cdots
\stackrel{\tilde M_1}{\Leftarrow} [\tilde{\bf z}_1,\tilde\lambda_1] \leftarrow [\tilde{\bf z}_0^\prime,\tilde\lambda_1] \stackrel{\tilde M_0}{\Leftarrow} [\tilde{\bf z}_0,\tilde\lambda_0]
\end{equation}
where $\tilde{\bf z}_{i}^\prime \equiv \tilde M_i(\tilde{\bf z}_{i})$, and the initial state $\tilde{\bf z}_0$ is sampled from $\rho_{eq}^B$.
The direction of the arrows indicates the progression of time. The probability density for obtaining a trajectory $\tilde \gamma=\{ \tilde {\bf z}_{0},\tilde {\bf z}_{1},\dots,\tilde {\bf z}_{N-1} \}$ is
\begin{equation}
\label{Map.RgammaP}
p[\tilde\gamma]= 
P_{\tilde \lambda_{N-1}}(\tilde {\bf z}_{N-1}|\tilde {\bf z}_{N-2}^\prime) ,
\cdots 
P_{\tilde \lambda_{2}}(\tilde {\bf z}_{2}|\tilde {\bf z}_{1}^\prime) \,
P_{\tilde \lambda_{1}}(\tilde {\bf z}_{1}|\tilde {\bf z}_{0}^\prime) \,
\rho_{eq}^B(\tilde{\bf z}_0) 
\end{equation}
with $d\tilde\gamma = d\tilde{\bf z}_0 \cdots d\tilde {\bf z}_{N-1}$.
Following Eq.~\ref{Map.work}, the work performed during this process is
\begin{equation}
\label{Map.Rwork}
\begin{split}
&W_R[\tilde \gamma]= \sum_{i=0}^{N-1} \left[ H_{\tilde \lambda_{i+1}}(\tilde {\bf z}_i^\prime)-H_{\tilde \lambda_{i}}(\tilde {\bf z}_i) - \beta^{-1}\ln \tilde J_i(\tilde {\bf z}_i) \right] ,
\end{split}
\end{equation}
where $\tilde J_i(\tilde{\bf z}) = \vert \partial\tilde{\bf z}^\prime/\partial\tilde{\bf z}\vert$ is the Jacobian for the mapping $\tilde M_i$.
Here and below we use the subscripts $F$ and $R$ to specify the forward and reverse processes, respectively.

We will now show that the work distributions corresponding to these two processes satisfy Crooks's fluctuation relation,~\cite{Crooksfluctuation,Crooks1998,Crooks99} namely
\begin{equation}
\frac{P_F(W)}{P_R(-W)}=e^{\beta (W-\Delta F)}
\end{equation}
where
\begin{equation}
P_F(W) = \int d\gamma \, p_F[\gamma] \, \delta \left( W - W_F[\gamma] \right)
\end{equation}
denotes the distribution of work values for the forward process, and $P_R(W)$ is similarly defined for the reverse process.

To establish this result, consider a {\it conjugate pair} of trajectories, $\gamma$ and $\gamma^*$, related by time-reversal.
Specifically, if $\gamma = \{ {\bf z}_0, \cdots {\bf z}_{N-1} \}_F$ is a trajectory generated during the forward process, that visits the sequence of microstates
\begin{equation}
\label{eq:forwardTrajectory}
	{\bf z}_0 \stackrel{M_0}{\Rightarrow} {\bf z}_0^{\prime} \rightarrow {\bf z}_1
	\stackrel{M_1}{\Rightarrow} {\bf z}_1^\prime \rightarrow \cdots
	\rightarrow {\bf z}_{N-1}
	\stackrel{M_{N-1}}{\Rightarrow} {\bf z}_{N-1}^\prime \quad ,
\end{equation}
then its conjugate twin, $\gamma^* = \{ {\bf z}_{N-1}^\prime, \cdots {\bf z}_0^\prime \}_R$, generated during the reverse process, visits the same microstates, in reverse order:
\begin{equation}
\label{eq:reverseTrajectory}
	{\bf z}_0 \stackrel{\tilde M_{N-1}}{\Leftarrow} {\bf z}_0^{\prime} \leftarrow {\bf z}_1
	\stackrel{\tilde M_{N-2}}{\Leftarrow} {\bf z}_1^\prime \leftarrow \cdots
	\leftarrow {\bf z}_{N-1}
	\stackrel{\tilde M_0}{\Leftarrow} {\bf z}_{N-1}^\prime
\end{equation}
that is $\tilde{\bf z}_i = {\bf z}_{N-1-i}^\prime$ and $\tilde{\bf z}_i^\prime = {\bf z}_{N-1-i}$ (see Eq.~\ref{Map.scheme.R}).
Note that the primary microstates of $\gamma$ are the secondary microstates of $\gamma^*$, and vice-versa, and the work function is odd under time-reversal:
\begin{equation}
\label{eq:W_odd}
W_F[\gamma]=-W_R[\gamma^*].
\end{equation}

We wish to evaluate the quantity 
\begin{equation} 
\label{flu1}
P_F(W) \, e^{-\beta (W-\Delta F)}= \int d\gamma\, p_F[\gamma] \, e^{-\beta (W_F[\gamma]-\Delta F)} \, \delta (W-W_F[\gamma])
\end{equation}
with $p_F[\gamma]$ given by Eq. \ref{Map.gammaP}.
To this end, we first decompose $W_F[\gamma]$ as follows:
\begin{equation}
\label{Wsplit}
W_F[\gamma] = \Delta E_F[\gamma] - Q_F[\gamma] - \beta^{-1} S_F[\gamma],
\end{equation}
where
\begin{subequations}
\label{eq:definitions}
\begin{eqnarray}
\Delta E_F[\gamma] &\equiv& H_{\lambda_N}({\bf z}_{N-1}^\prime)-H_{\lambda_0}({\bf z}_0) \\
Q_F[\gamma] &\equiv& \sum_{i=1}^{N-1} \left[ H_{\lambda_{i}}({\bf z}_{i})-H_{\lambda_i}({\bf z}_{i-1}^\prime) \right] \\
S_F[\gamma] &\equiv& \sum_{i=0}^{N-1} \ln J_{\lambda_{i}}({\bf z}_i)
= \ln \prod_{i=0}^{N-1} \left\vert \frac{\partial {\bf z}_i^\prime}{\partial{\bf z}_i} \right\vert = \ln \left\vert \frac{\partial \gamma^*} {\partial \gamma} \right\vert
\end{eqnarray}
\end{subequations}
Here $\Delta E_F[\gamma]$ is the total change in the energy of the system as it evolves along the trajectory $\gamma$, $Q_F[\gamma]$ can be interpreted as the heat transfered to the system from the reservoir~\cite{miller:7035}, and $S_F[\gamma]$ is an entropy-like term, which arises because the mappings $M_i$ need not preserve volume.
The quantities defined in Eq.~\ref{eq:definitions} satisfy the properties
\begin{subequations}
\begin{eqnarray}
\label{heatflip} 
P_{\lambda_{N-1}}({\bf z}_{N-1}|{\bf z}_{N-2}^\prime)\cdots P_{\lambda_1}({\bf z}_1|{\bf z}_0^\prime) 
&=&
P_{\lambda_{N-1}}({\bf z}_{N-2}^\prime|{\bf z}_{N-1})\cdots P_{\lambda_1}({\bf z}_0^\prime|{\bf z}_1)\, e^{-\beta Q_F[\gamma]} \\
\label{eflip}
\rho_{eq}^{\lambda_0}({\bf z}_0) 
&=&
\rho_{eq}^{\lambda_{N}}({\bf z}_{N-1}^\prime) \, e^{\beta (\Delta E_F[\gamma]-\Delta F)} 
\end{eqnarray}
\end{subequations}
where we have used Eqs.~\ref{equilibrium} and \ref{D.Balance}.
These properties then give us
\begin{equation}
\label{flu2}
\begin{split}
p_F[\gamma] &= 
P_{\lambda_{N-1}}({\bf z}_{N-1} | {\bf z}_{N-2}^\prime) 
\cdots 
P_{\lambda_1}({\bf z}_1 \vert {\bf z}_0^\prime) \,
\rho_{eq}^{\lambda_0}({\bf z}_0) \,
\\
&=
P_{\lambda_{N-1}}({\bf z}_{N-2}^\prime|{\bf z}_{N-1}) \cdots P_{\lambda_1}({\bf z}_0^\prime|{\bf z}_1) \, e^{-\beta Q_F[\gamma]}\\
&\qquad \times \,
\rho_{eq}^{\lambda_N}({\bf z}_{N-1}^\prime) \, e^{\beta (\Delta E_F[\gamma] - \Delta F)}
\ \\
&=
p_R[\gamma^*] \, e^{\beta(W_F[\gamma]-\Delta F)} \, e^{S_F[\gamma]} 
\end{split}
\end{equation}
hence
\begin{equation}
p_F[\gamma] \, e^{-\beta (W_F[\gamma]-\Delta F)} =
p_R[\gamma^*] \, \left\vert \frac{\partial \gamma^*}{\partial \gamma} \right\vert
\end{equation}
Substituting this result into the integrand on the right side of Eq.~\ref{flu1}, then changing the variables of integration from $d\gamma$ to $d\gamma^*$, and invoking Eq.~\ref{eq:W_odd}, we finally arrive at the result we set out to establish:
\begin{equation}
\label{fr:ratio}
P_F(W) \, e^{-\beta (W-\Delta F)}=P_R(-W)
\end{equation}

Eq. \ref{fr:ratio} in turn implies that the average of any function $f(W)$ over work values generated in the forward process, can be related to an average over work values obtained in the reverse process:~\cite{Crooksfluctuation} 
\begin{equation}
\label{Crooks:FT}
\frac{\langle f(W) \rangle_F}{\langle f(-W) e^{-\beta W}\rangle_R }=e^{-\beta \Delta F}
\end{equation}
In principle, this result can be used with any $f(W)$ to estimate $\Delta F$.
The problem of determining the optimal choice of $f(W)$ was solved by Bennett in the context of equilibrium sampling,~\cite{Bennett} and this solution can be applied directly to the nonequilibrium setting.~\cite{Crooksfluctuation,Bennett:Likelihood}
Specifically, if we have $n_F$ work values from the forward simulation, and $n_R$ work values from the reverse simulation, then the optimal choice is
\begin{equation}
f(W)= \frac{1}{1+\exp (\beta W +\beta K)}
\end{equation}
where $K=-\Delta F+\beta^{-1} \ln (n_F/n_R)$.
The value of $\Delta F$ is then estimated by recursively solving the equation,
\begin{equation}
\label{Bennett:FT}
e^{-\beta \Delta F}=\frac{\langle 1/(1+e^{\beta (W +K)}) \rangle_F}{\langle 1/(1+e^{\beta (W -K)} )\rangle_R }e^{\beta K}
\end{equation}
as described in detail in Ref.~\cite{Bennett}.
This procedure for estimating $\Delta F$ is known as {\it Bennett's Acceptance Ratio }method (BAR).

\section{Computational efficiency and figures of merit}
\label{Merit}
While Eqs. \ref{Map.discreteF} and \ref{Bennett:FT} are valid for any set of invertible mapping functions, $\{M_i\}$, the efficiency of using escorted simulations to estimate $\Delta F$ depends strongly on the choice of these functions.
Since the convergence of exponential averages such as Eq.~\ref{Map.discreteF} deteriorates rapidly with dissipation~\cite{Kofke,CJ:Rareevents,Gore}, which in turn correlates with the lag illustrated in Fig.~(\ref{lag.sketch}), it is reasonable to speculate that a choice of mappings that decreases the lag will improve the convergence of estimator (Eq. \ref{Map.discreteF}).

To pursue this idea, let us first consider the extreme case of a set of mapping functions $\{M^*_i\}$ that entirely eliminates the lag.
By this we mean the following:
for an ensemble of trajectories generated using Eq.~\ref{Map.scheme}, with ${\bf z}_0$ sampled from $p_{eq}^A({\bf z}_0)$, the subsequent microstates ${\bf z}_i$ are distributed according to $p_{eq}^{\lambda_i}({\bf z}_i)$, for all $1\leq i<N$.
That is, the shaded and unshaded ovals coincide in Fig.~(\ref{lag.sketch}).
This occurs if under the bijective mapping $M^*_i:{\bf z}\rightarrow{\bf z}^\prime$, the equilibrium distribution $\rho^{\lambda_i}_{eq}({\bf z})$ transforms to the distribution $\rho^{\lambda_{i+1}}_{eq}({\bf z}^\prime)$~\cite{CJ:Targeting}, in other words
\begin{equation}
\label{perfectM}
\rho^{\lambda_{i+1}}_{eq}({\bf z}^\prime)=\frac{\rho^{\lambda_i}_{eq}({\bf z})}{J_{\lambda_i}^*({\bf z})}
\end{equation}
[Under a bijective map $M:{\bf x}\rightarrow{\bf y}$, a distribution $f({\bf x})$ is transformed to the distribution $\eta({\bf y})=f({\bf x})/J({\bf x})$, where $J({\bf x}) = \vert \partial{\bf y}/\partial{\bf x} \vert$.]
When all the $M^*_{\lambda_i}$'s satisfy this condition, we will say that the set of mappings is {\it perfect}.
Using $\rho_{eq}^\lambda=e^{\beta (F_\lambda-H_\lambda)}$, and taking the logarithm of both sides of Eq. \ref{perfectM}, we obtain (for a perfect set of mappings)
\begin{equation}
\label{perfect}
 \delta W_i \equiv H_{\lambda_{i+1}}({\bf z}^\prime)-H_{\lambda_{i}}({\bf z})-\beta^{-1} \ln J_{\lambda_i}^*({\bf z})=F_{\lambda_{i+1}}-F_{\lambda_i} ,
\end{equation} 
hence $W[\gamma]=\Delta F$ {\it for every trajectory} $\gamma$ (Eq.~\ref{Map.work}).
Thus for a perfect set of mappings we have $P_F(W) = \delta(W-\Delta F)$, and Eq. \ref{Map.discreteF} provides a zero-variance estimate of the free energy difference.
It is straightforward to show that if the $M^*_i$'s form a set of perfect mappings for the forward process, then the $\tilde M^*_i$'s form a set of perfect mappings for the reverse process, and $P_R(W) = \delta(W+\Delta F)$.

The considerations of the previous paragraph support the idea that reducing lag improves convergence.
While we generally cannot expect to be able to construct a {\it perfect} set of mapping functions (this is likely to be far more difficult than the original problem of estimating $\Delta F$!~\cite{escorted}), in many cases it might be possible to use either intuition or prior information about a system to construct a set of $M_i$'s that reduce the lag substantially. 
In such cases the dissipation accompanying the escorted simulations is less than that for the unescorted simulations, leading to improved convergence of the free energy estimate.

As an example of a strategy that can be used to construct good mappings, consider a system of identical, mutually interacting particles, in an external potential $U_\lambda({\bf r})$:
\begin{equation}
\label{fullH}
H_\lambda({\bf z}) = \sum_k U_\lambda({\bf r}_k) +  \sum_{k<l} V({\bf r}_k,{\bf r}_l)
\end{equation}
The probability distribution of a single, tagged particle is then given by the single-particle density
\begin{equation}
\rho_\lambda^{(1)}({\bf r}) =
\frac{1}{Z_\lambda}
\int d{\bf z} \, \delta[{\bf r}_k({\bf z}) - {\bf r}] \, e^{-\beta H_\lambda({\bf z})}
\end{equation}
where ${\bf r}_k({\bf z})$ specifies the coordinates of the tagged particle as a function of the microstate ${\bf z}$.
Now consider a reference system of non-interacting particles, described by a Hamiltonian
\begin{equation}
\label{mimicH}
\bar{H}_\lambda({\bf z}) = \sum_k \bar{U}_\lambda({\bf r}_k)
\end{equation}
with a similarly defined single-particle density $\bar{\rho}_\lambda^{(1)}({\bf r})$;
and imagine that $\bar{U}_\lambda$ is chosen so that these single-particle densities are identical or nearly identical: $\rho_\lambda^{(1)}({\bf r}) \approx \bar{\rho}_\lambda^{(1)}({\bf r})$.
In this case a set of mappings $\{ M_i \}$ that are perfect or near-perfect for the reference system ($\bar H_\lambda$), might be quite effective in reducing lag in the original system ($H_\lambda$).
We will illustrate this mean-field-like approach in Section \ref{subsec:dipoleFluid}, and we note that a similar strategy was explored by Hahn and Then in the context of targeted free energy perturbation~\cite{BijectiveTFEP}.

It will be useful to develop a figure of merit, allowing us to compare the efficiency of our method for different sets of mappings. One approach would be simply to compare the error bars associated with the statistical fluctuations in the respective free energy estimates. Unfortunately, estimates of $\Delta F$ obtained from convex nonlinear averages such as the one obtained from Eq. ~\ref{Map.discreteF}, are systematically biased for any finite number of realizations~\cite{Gore,Zuckerman02}. This bias can be large, and as a result the statistical error bars by themselves might not be sufficiently reliable to quantify the efficiency of the mapping. In the following paragraphs we discuss alternative figures of merit.

We begin by noting that when the unidirectional estimator, Eq.~\ref{Map.discreteF}, is used in conjunction with simulations of the forward process, then the number of realizations ($N_s$) required to obtain a reliable estimate of $\Delta F$ is roughly given by~\cite{CJ:Rareevents,Kofke} 
\begin{equation} 
\label{NUnidirectional}
N_s\sim e^{\beta(\langle W \rangle_R+\Delta F)}
\end{equation}
where  $\langle W\rangle_R+\Delta F$ is the dissipation accompanying the reverse process.
While this provides some intuition for the convergence of Eq.~\ref{Map.discreteF}, its usefulness as a figure of merit is somewhat limited as it requires simulations of both the forward and the reverse processes, and in that case we are better off using a bidirectional estimator such as Eq.~\ref{Bennett:FT}.

When we do have simulations of both processes, then an easily computed figure of merit is the hysteresis, $\langle W_{diss}\rangle _F+\langle W_{diss}\rangle_R=\langle W \rangle_F+\langle W\rangle_R$. 
The value of this quantity is zero if the mappings are perfect, otherwise it is positive.
It is interesting to note that the hysteresis can be related to an information-theoretic measure of overlap between the forward and reverse work distributions $P_F(W)$ and $P_R(-W)$:~\cite{PhysRevLett.101.090602}
\begin{equation}
\label{Relativeentropy}
D[P_F || P_R ]+D[P_R ||P_F ]=
\beta(\langle W \rangle_F+\langle W \rangle_R) .
\end{equation}
Here $D[p||q]\equiv \int p\ln(p/q) \ge 0$ denotes the relative entropy between the distributions $p$ and $q$, and the symmetrized quantity $D[p||q]+D[q||p]$ (also known as the {\it Jeffreys divergence}~\cite{Cover2006}) provides a measure of the difference, or more precisely the lack of overlap, between the distributions.
The right side of Eq.~\ref{Relativeentropy} can be estimated from a modest sample of forward and reverse simulations.
If a set of mappings reduces the hysteresis, $\langle W \rangle_F+\langle W \rangle_R$, then this indicates increased overlap between the work distributions, and therefore improved convergence~\cite{CJ:Rareevents}.

When $n_F=n_R=N_s \gg 1$, the mean square error of the Bennett estimator is~\cite{Bennett:Likelihood,BijectiveTFEP,Bennett,HahnThen2010}
\begin{equation}
\label{MSEBAR}
\langle ( F^{est}_{BAR}-\Delta F)^2\rangle=\frac{2}{\beta^2N_s} \left( \frac{1}{2 C}-1 \right) .
\end{equation}
Here $F^{est}_{BAR}$ denotes the estimate of $\Delta F$ obtained from Eq. \ref{Bennett:FT}, and 
\begin{equation}
\label{Cdefine}
C \equiv \int dW \frac{ P_F(W)P_R(-W)}{P_F(W)+P_R(-W)}= \left\langle \frac{1}{1+\exp [\beta(W -\Delta F)]} \right\rangle_F=
\left\langle \frac{1}{1+\exp [\beta(W +\Delta F)]} \right\rangle_R
\end{equation}
(This result can be generalized to the case $n_F \ne n_R$~\cite{BijectiveTFEP}.)
As discussed by Bennett~\cite{Bennett} and Hahn and Then~\cite{BijectiveTFEP,HahnThen2010}, the value of $C$ measures the overlap between $P_F(W)$ and $P_R(-W)$,
and provides a rough figure of merit for the Bennett estimator.
When lag is eliminated and the two distributions coincide, then $C$ attains its maximum value, $C=1/2$, whereas when there is poor overlap, $C\approx 0$. 
Thus we expect that the higher the value of the overlap function $C$, the smaller the number of realizations $N_s$ required to estimate $\Delta F$ from Eq.~\ref{Bennett:FT} with a prescribed accuracy.
Indeed, Eq.~\ref{MSEBAR} suggests a lower bound on the number of realizations needed to achieve a mean square error less than $\beta^{-2}$: $N_s> 1/ C$.
Note that since $C$ is an ensemble average (Eq.~\ref{Cdefine}), it can readily be estimated from available simulation data.

In the Appendix, we derive an upper bound on the number of realizations needed to obtain a reliable estimate of $\Delta F$ using Bennett's method, $N_s$ (Eq. \ref{Nbennett}). Combining these bounds gives us
\begin{equation}
\label{Cbound}
\frac{1}{C} < N_s<  \frac{1}{C^2} 
\end{equation}
While Eq. \ref{Cbound} cannot be used to obtain a good estimate for $N_s$~\footnote{For $C<<1$, the upper and lower bounds in Eq. \ref {Cbound} can be
  orders of magnitude apart. Nevertheless, Eq. \ref {Cbound} can serve as a
  good consistency check for the quality of the estimates. For example an
  estimate of $\Delta F$ using Bennett's method from a data set of size
  $N_s\sim 10^6$ is reliable if $C\sim 0.001$.}, it does allow us to argue heuristically that whenever a set of mappings succeeds in increasing the value of $C$, the convergence of the Bennett estimator is improved.
We will illustrate this point in the following section.

\section{Examples}
\label{Examples}
\subsection{Cavity Expansion}

\begin{figure}[tbp]
 \includegraphics[scale=0.58,angle=0]{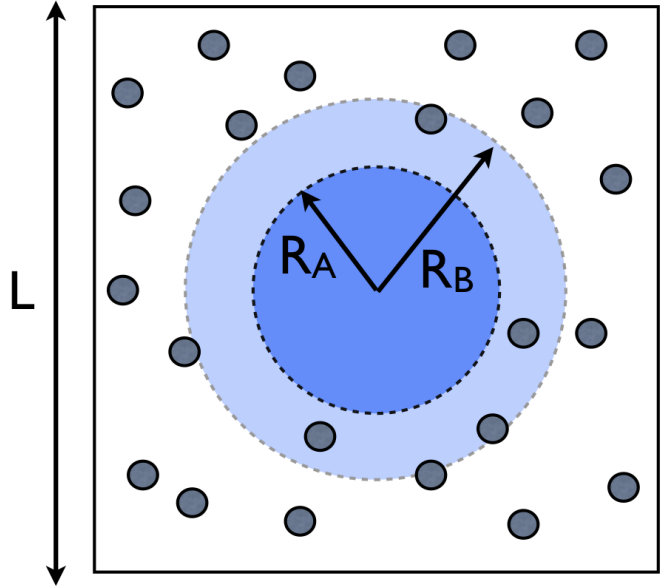} 
 \caption{A schematic of the cavity expansion problem}
 \label{Fig:Cavity}
 \end{figure}

As a first example, we estimate the free energy cost associated with growing a hard-sphere solute in a fluid. Consider a system composed of $n_p$ point particles inside a cubic container of volume $L^3$, centered at the origin with periodic boundaries. 
The particles are excluded from a spherical region of radius $R$, also centered at the origin. The particles interact with one another via the WCA pairwise interaction potential~\cite{Frenkel} which is denoted by $V({\bf r}_k,{\bf r}_l)$. The energy of the system at a microstate ${\bf z}= ( {\bf r}_1,{\bf r}_2,\dots,{\bf r}_{n_p} )$ is given by 
\begin{equation}
\label{Energy}
H_R({\bf z})= \Theta({\bf z},R)+\sum_{k=1}^{n_p-1}\sum_{l>k}^{n_p} V({\bf r}_k,{\bf r}_l)
\end{equation}
where $ \Theta({\bf z},R)=0$ whenever $|{\bf  r}_k|>R$ for all $k=1, \cdots n_p$, that is when there are no particles inside the spherical cavity; and $ \Theta({\bf z},R)=\infty$ otherwise. The function $\Theta({\bf z},R)$ ensures that particles are excluded from the spherical region around the origin. We wish to compute the free energy cost, $\Delta F$, associated with increasing the radius of the cavity from $R_A$ to $R_B$ (See Fig. \ref{Fig:Cavity}).
 
A hypothetical estimate of $\Delta F$ using {\it unescorted} nonequilibrium simulations (Eq.~\ref{scheme}) involves ``growing out'' the spherical cavity in discrete increments, as follows.
Starting with a microstate ${\bf z}_0$ sampled from equilibrium at $R=R_A$, the radius of the sphere is increased by an amount $\delta R_0$.
If all $n_p$ fluid particles remain outside the enlarged sphere, then $\delta W_0=0$;
but if one or more particles now finds itself inside the sphere ($r_k < R_A+\delta R_0$) then $\delta W_0=\infty$.
One or more Monte Carlo steps are then taken, after which the radius is again increased by some amount, $\delta R_1$, and $\delta W_1$ is determined in the same fashion as $\delta W_0$.
In principle this continues until the radius of the sphere is $R_B$, and then the work is tallied for the entire trajectory: $W = \sum_i \delta W_i$.
In practice the trajectory can be terminated as soon as $\delta W_i = \infty$ at some step $i$, since this implies $W=\infty$.
For this procedure, Eq.~\ref{NEW} can be rewritten as
\begin{equation}
\label{eq:PdF}
P = e^{-\beta\Delta F} ,
\end{equation}
where $P$ is the probability of generating a trajectory for which $W=0$;
that is, a trajectory in which the sphere is successfully grown out to radius $R_B$, without overtaking any fluid particles along the way.
The quantity $P$ is estimated directly, by generating a number of trajectories and counting the ``successes'' ($W=0$).
For a sufficiently dense fluid, however, a successful trajectory is a rare event ($P\ll 1$), and this approach converges poorly.
Note also that this approach does not give the correct free energy difference in the reverse case of a shrinking sphere (from $R=R_B$ to $R=R_A$), since $W=0$ for every trajectory in that situation.

For the hypothetical procedure just described, Eq.~\ref{eq:PdF} implies that the probability to generate a successful trajectory does not depend on the number of increments used to grow the cavity from $R_A$ to $R_B$.
Therefore the most computationally efficient implementation is to grow the sphere out in a single step, which corresponds to the free energy perturbation method (FEP)~\cite{Chipot2007,Frenkel}.
In this case $P$ is just the probability to observe no particles in the region $R_A < r < R_B$, for an equilibrium simulation at cavity radius $R_A$.

To improve convergence by means of escorted simulations (Eq.~\ref{Map.scheme}), we constructed mapping functions $M_i$ that move the fluid particles out of the way of the growing sphere, to prevent infinite values of $\delta W_i$.
Specifically, as the cavity radius $R$ is increased from $R_i$ to $R_{i+1}$, the location of the $n^{th}$ particle, ${\bf r}_n$, is mapped to ${\bf r}_n^\prime=m_i({\bf r}_n)$, where~\cite{CJ:Targeting} 
\begin{equation}
    \label{Mapping}
    m_i({\bf r}_n) =
      \left [1+\frac{(R_{i+1}^3-R_i^3)(L^3-8r_n^3)}{(L^3-8R_i^3)r_n^3}\right ]^{1/3}{\bf r}_n  \qquad \text{if $r_n \le L/2$}
\end{equation}
and $m_i({\bf r}_n)={\bf r}_n$ if $r_n>L/2$.
The notation $m_i: {\bf r}_n \rightarrow {\bf r}_n^\prime$ denotes a single-particle mapping;
the full mapping $M_i:{\bf z} \rightarrow {\bf z}^\prime$ is obtained by applying $m_i$ to all $n_p$ fluid particles.
To picture the effect of this mapping, let ${\cal S}_i$ denote the region of space defined by the conditions $R_i \le r \le L/2$,
that is a spherical shell of inner radius $R_i$ and outer radius $L/2$ (just touching the sides of the cubic container).
Under the mapping
$m_i: {\bf r}\rightarrow {\bf r}^\prime$,
the shell ${\cal S}_i$ is compressed uniformly onto the shell ${\cal S}_{i+1}$, leaving the eight corners of the box $r>L/2$ untouched. \footnote{
An even better mapping would uniformly compress the entire region $r>R_i$, including the eight corners, onto the region $r>R_{i+1}$.
However, due to the geometric mismatch between the spherical inner surface and cubic outer surface of these regions, such a mapping is not represented by a simple formula such as Eq.~\ref{Mapping}, and would need to be constructed numerically.
}
\begin{figure}[tbp]
  \includegraphics[scale=0.5,angle=-90]{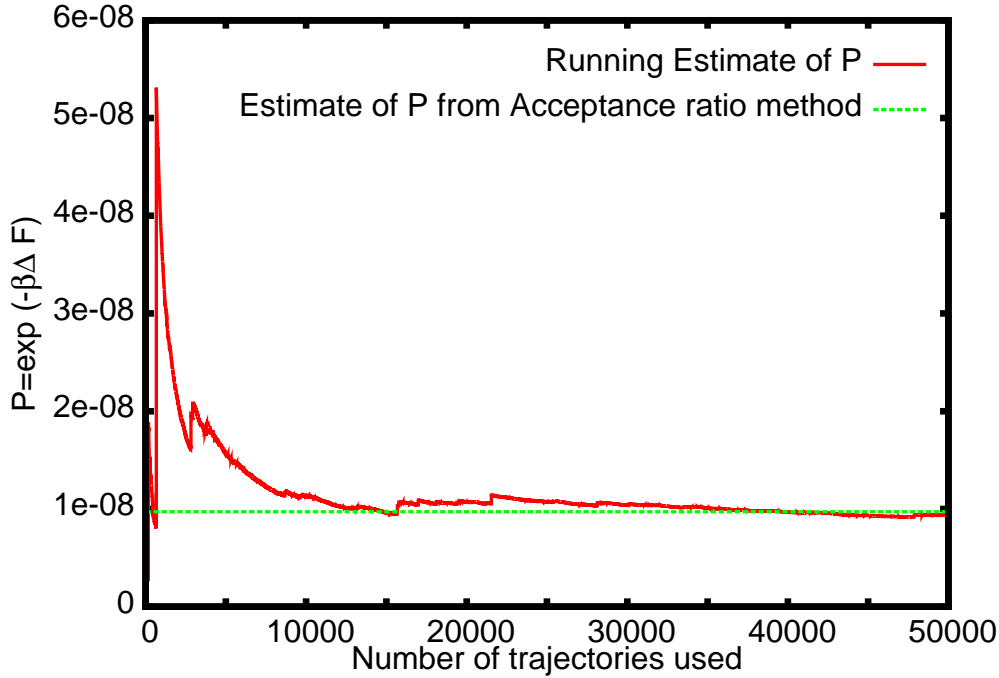}
   \caption{Running estimate of $P=\exp(-\beta \Delta F)$ from escorted free energy simulations, plotted as a function of the number of trajectories used to obtain the estimate.
    }   
 \label{Fig:WCA.run}
   \end{figure}
In this manner, the particles that would otherwise have found themselves inside the enlarged sphere are pushed outside of it, resulting in a finite contribution to the work (Eq.~\ref{Map.work}),
\begin{equation}
\label{eq:dW_cavity}
\delta W_i = \sum_{k=1}^{n_p-1}\sum_{l>k}^{n_p} \left[ V({\bf r}_k^\prime,{\bf r}_l^\prime) - V({\bf r}_k,{\bf r}_l) \right]
- n_0 \beta^{-1} \ln \gamma
\end{equation}
where $n_0 = n_0({\bf z})$ is the number of particles found within the shell $R_i \le r \le L/2$ (before the mapping is applied), and
$\gamma = (L^3-8R_{i+1}^3)/(L^3-8R_i^3) < 1$ is the ratio of shell volumes, $\vert {\cal S}_{i+1} \vert / \vert {\cal S}_i \vert$.
The first term on the right side of Eq.~\ref{eq:dW_cavity} gives the net change in the energy of the system associated with the escorted switch
$[{\bf z}_i,R_i]\stackrel{M_i}{\Rightarrow}[{\bf z}_i^{\prime},R_{i+1}]$, while the second is the Jacobian term $-\beta^{-1} \ln J_i({\bf z}_i)$.

Unlike the unescorted approach or free energy perturbation, the escorted approach with the mapping given by Eq. \ref{Mapping} is applicable in both the forward (growing spherical cavity) and reverse (shrinking cavity) directions.
In the reverse direction, as the solute radius is decreased from $R_{i+1}$ to $R_i$, the shell ${\cal S}_{i+1}$ is uniformly expanded onto the shell ${\cal S}_i$.
The corresponding increment in work is given by a formula similar to Eq.~\ref{eq:dW_cavity}.
As a result, one can combine work values from forward and reverse escorted simulations using Bennett's Acceptance Ratio (BAR), Eq. \ref{Bennett:FT}. 
\begin{table}
\begin{tabular}{|c|c|}
\hline
$\langle W\rangle_F$&22.288$\pm$ 0.012\\
\hline
$\langle W\rangle_R$&-14.458$\pm$ 0.013\\
\hline
$\langle W \rangle_F+\langle W\rangle_R$&7.830$\pm$ 0.018\\
\hline
$\Delta F^{est}_{F}$&$18.487\pm 0.085$\\
\hline
$\Delta F^{est}_{R}$&$-18.334\pm 0.078$\\
\hline

$\Delta F^{est}_{BAR} $& $18.456\pm 0.011$\\
\hline
C &$0.120 \pm 0.001$\\
\hline
\end{tabular}
\caption{Estimates and figures of merit. Here $\Delta F^{est}_F$ denotes the estimate of $\Delta F\equiv F_B-F_A$ from the forward process ($R_A \rightarrow R_B$) and $\Delta F^{est}_R$ denotes the estimate of $-\Delta F$ from the reverse process ($R_A \leftarrow R_B$). $\Delta F^{est}_{BAR}$ denotes the estimate of $\Delta F$ obtained from Bennett's Acceptance Ratio method.
} 
\label {Table:0}
\end{table}

We have performed both forward and reverse simulations of this system using $n_p = 1000$ WCA particles, with $L=10.42 \sigma$, $R_A=2.0 \sigma$, $R_B=2.05 \sigma$, and $T^* \equiv k_B T/\epsilon=1$, where the WCA parameters $\sigma$ and $\epsilon$ set the units of length and energy, respectively.
Minimum image convention and periodic boundary conditions were used~\cite{Frenkel}.

 Fig \ref{Fig:WCA.run} shows a running estimate of $P=\exp(-\beta \Delta F)$ obtained from escorted simulations in which the solute radius was switched from $R_A$ to $R_B$ in $N=10$ steps, with each increment in $R$ alternating with one Monte Carlo sweep.
Using a total of $N_s=50000$ independent escorted trajectories, estimates of $\Delta F$ and the figures of merit were obtained, and are summarized in Table \ref{Table:0} (The value of $C$ and $\Delta F^{est}_{BAR}$ were estimated using $n_F=n_R=N_s=50000$ trajectories). Statistical error bars were computed using the bootstrap method~\cite{Bootstrap}.
While an analytical expression for $\Delta F$ is not available for this example, the agreement between the estimates obtained by growing the solute ($F$), shrinking it ($R$), and applying BAR gives us confidence in the result, $\Delta F \approx 18.4~k_BT$.

\begin{figure}[tbp]      
             \includegraphics[scale=0.5,angle=-90]{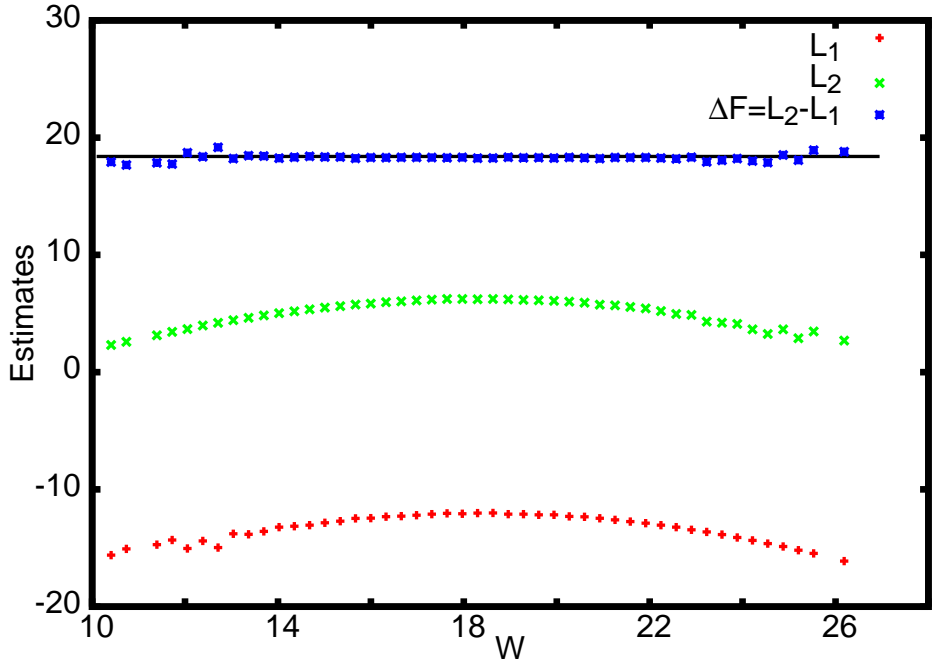} 
              \caption{Graphical verification of the fluctuation theorem and estimation of $\Delta F$.
              The horizontal line indicates the estimate of $\Delta F$ obtained from the acceptance ratio method (Table~\ref{Table:0}).
 		}             
 \label{WFXprod_ben4}   
               \end{figure}

As an additional consistency check, in Fig.~\ref{WFXprod_ben4} we verify that the escorted simulations satisfy the fluctuation theorem Eq. \ref{fr:ratio}. We do this by following steps analogous to those in Section III of Ref.~\cite{Bennett} to obtain a restatement of Eq. \ref{fr:ratio},
\begin{equation}
\label{Bennett:Histogram}
L_2(W)-L_1(W) \equiv
\left[ \ln P_R(-W)+\beta \frac{W}{2} \right] - \left[ \ln P_F(W)-\beta \frac{W}{2} \right]=\beta \Delta F
\end{equation}
In Fig \ref{WFXprod_ben4} we plot $L_1$, $L_2$, and $L_2-L_1$ as functions of $W$.
The flatness of the difference $L_2-L_1$ over the region for which we have good statistics is in agreement with Eq.~\ref{Bennett:Histogram}, and provides a useful and stringent consistency check~\cite{Frenkel,GoodPractices2010}, which gives us further confidence in our estimates.

While the highly accurate estimates listed in Table~\ref{Table:0} were generated using $N_s=50000$ escorted trajectories, we found that we were able to obtain estimates of $\Delta F$ with error bars around $1\ k_BT$ using only $N_s=100$ realizations for the unidirectional estimators, and $N_s=10$ realizations for the bidirectional estimator (data not shown).

To compare the escorted method with unescorted free energy perturbation (FEP), we first sampled $N_s=100000$ independent configurations from the canonical ensemble with cavity radius $R=R_A$, by generating a single, long equilibrium Monte Carlo trajectory and sampling one configuration per 10 Monte Carlo sweeps.
This involved a total computational time approximately equal to that of generating 50000 escorted trajectories.
Among these $10^5$ configurations we did not observe a single one in which the region $R_A \leq r \leq R_B$ was spontaneously devoid of particles ($W=0$), in other words we were unable to obtain an estimate of $\Delta F$ using free energy perturbation. 
This is consistent with the result $P \approx e^{-18.4} \approx 10^{-8}$ (Fig.~\ref{Fig:WCA.run}, Table \ref{Table:0}), which suggests that roughly $10^8$ independent configurations are needed to observe one for which $W=0$.

For a more efficient implementation of FEP, we divided the interval $[R_A , R_B]$ into ten stages (sub-intervals), and then used FEP to estimate the free energy change for each stage, keeping the total computational time fixed.
This provided a final estimate of $\Delta F$ with error bars comparable to those of the unidirectional escorted estimators in Table \ref{Table:0}, but still considerably larger than those of the bidirectional estimates (data not shown).~\footnote{
Of course, even after dividing the problem into stages, one can apply escorting by separately treating each stage as a switching simulation with one step, $N=1$, and using the mappings given by Eq.~\ref{Mapping}.
We found that this further reduces the error bars by nearly a factor of six.
}

\subsection{Dipole Fluid} 
\label{subsec:dipoleFluid}
As our second example, we consider $n_p$ point Lennard-Jones dipoles in a cubic container of size $L$ with periodic boundaries, and we compute the free energy cost associated with introducing a uniform electric field in the container. The energy of the system in an external electric field ${\bf E}=E \hat {\bf e}_z $, where $\hat {\bf e}_z$ denotes a unit vector along the z-axis, is given by 

\begin{equation}
\begin{split}
\label{dipoleenergy}
H_{{ E},\gamma}({\bf z})=-\sum_{k=1}^{n_p} {\bf p}_k\cdot{\bf E}&+\sum_{k=1}^{n_p-1}\sum_{l>k}^{n_p} V_{LJ}({\bf r}_k,{\bf r}_l)-\gamma \frac{{\bf p}_k \cdot {\bf p}_l}{{|{\bf r}_k-{\bf r}_l|}^4}\\
\end{split}
\end{equation}
where ${\bf z}=\{ {\bf r}_1,{\bf p}_1,\dots {\bf r}_{n_p}, {\bf p}_{n_p}\}$, ${\bf p}_k$ denotes the dipole moment vector of the $k^{th}$ particle, and $V_{LJ}({\bf r}_k,{\bf r}_l)$ denotes the Lennard-Jones pairwise interaction potential. The parameter $\gamma$ controls the strength of the dipole-dipole interaction. We set $\vert {\bf p}_k \vert=1$ for all $k$. In spherical polar coordinates, ${\bf p}_k=(1,\theta_k,\phi_k)$, and the measure on ${\bf z}$ space is hence $d{\bf z}=\Pi_{k=1}^{n_p} d {\bf r}_k d\cos(\theta_k) d\phi_k$.

Taking the electric field to be the external parameter, we wish to compute the free energy difference between the ensembles corresponding to ${E}=0$ and ${ E}=E_f$ at some temperature $\beta^{-1}$ by performing nonequilibrium switching simulations. Our first task is to construct a mapping function that escorts the system along a near equilibrium path as $E$ is switched. Following Eq. \ref{mimicH}, we consider the energy function $\bar H_{{ E}}({\bf z})\equiv H_{{ E},0}({\bf z})$ (i.e. $\gamma =0$ in Eq. \ref{dipoleenergy}), which describes a system of non-interacting Lennard-Jones dipoles in a field of strength ${ E}$. The change in free energy as the field is switched from $E_{i}$ to $E_{i+1}$ can be solved analytically and is given by 
\begin{equation}
\label{exactfree-energy}
\bar F_{E_{i+1}}-\bar F_{E_i}=-n_p \frac{1}{\beta} \ln \left [ \frac{\sinh(\beta E_{i+1})}{\sinh(\beta E_i)} \frac{E_i}{ E_{i+1}} \right ]
\end{equation}
We now use this result to solve for a perfect set of mappings for this system of non-interacting dipoles.

Let $ m_i:\zeta\equiv \cos(\theta)\rightarrow \zeta^\prime$ denote a mapping that acts on the $\zeta=\cos(\theta)$ degree of freedom of a dipole when the external field is switched from $E_i$ to $E_{i+1}$. The full mapping $M_i$ is obtained by applying the mapping $m_i$ to all $n_p$ particles. We look for the perfect mapping $M_i$ that  transforms the canonical distribution corresponding to $\bar H_{{ E}_{i}}({\bf z})$ to the canonical distribution corresponding to $\bar H_{{ E}_{i+1}}({\bf z}^\prime)$. The following equation for the perfect single particle mapping $m_i$ can be obtained from Eq. \ref{perfect} by using Eqs.~\ref{dipoleenergy} and \ref{exactfree-energy} and by noting that ${\bf p}_k \cdot {\bf E}= E \zeta_k$:
\begin{equation}
\label{perfect1}
{ E_{i+1} m_i(\zeta)-  E_{i} \zeta}- \frac{1}{\beta} \ln \frac{ d m_i(\zeta)}{d \zeta}=-\frac{1}{\beta}\ln  \frac{\sinh(\beta E_{i+1})}{\sinh(\beta E_i)}\frac {E_i}{E_{i+1}}
\end{equation}
\begin{figure}[tb]
 \includegraphics[scale=0.5,angle=-90]{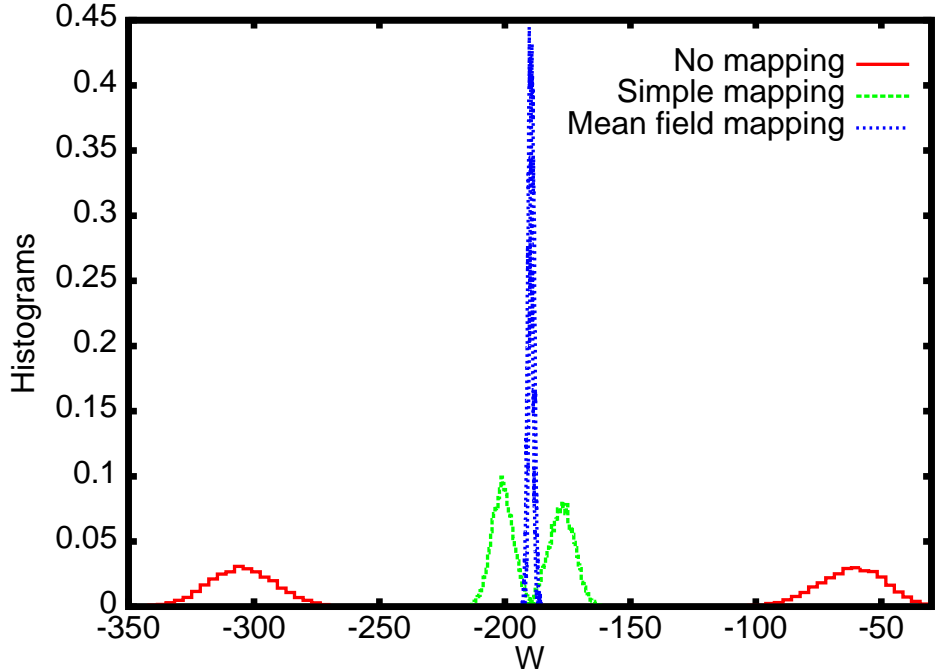}
 \caption{Work histograms obtained from forward and reverse simulations performed at $\gamma=0.1$.
 The degree of overlap between $P_F(W)$ and $P_R(-W)$ provides an indication of the efficiency of the free energy estimate.
 For unescorted simulations (no mapping) we see no overlap, reflecting considerable dissipation and poor efficiency
 (Table \ref{Tab:2}).
 With the mapping given by Eq.~\ref{PerfectMdipole} the overlap is much improved, and with the mean field mapping, Eq.~\ref{MeanFMdipole} the forward and reverse distributions are nearly identical.}
  \label{data2.his}
 \end{figure}
This differential equation has the solution
\begin{equation}
\label{PerfectMdipole}
m_i(\zeta)=\frac{1}{\beta E_{i+1}} \ln\left [\frac{\sinh(\beta E_{i+1})}{\sinh(\beta E_i)}(e^{\beta E_i \zeta}-e^{\beta E_i})+e^{\beta E_{i+1}}\right ]
\end{equation}
While Eq. \ref{PerfectMdipole} is a perfect mapping only when there are no dipole-dipole interactions ($\gamma=0$) we expect this mapping to work reasonably well for small values of $\gamma$.
We will use the term {\it simple mapping} in reference to Eq.~\ref{PerfectMdipole}.

We also constructed a set of mapping functions using mean field~\cite{Chandler} arguments as follows. In the absence of long range order, mean field theory suggests that the interacting dipole-fluid system ($\gamma \neq 0$) in an electric field of strength ${E}$ can be approximated by a system of non-interacting dipoles ($\gamma = 0$) in an effective field of strength $E^\prime$. We obtained approximate values for this effective electric field by first numerically evaluating the single-dipole distribution $P(\zeta) \, ,\zeta=\cos(\theta)$, at $E=E_f$. The thermal distribution of $\zeta$ for a non-interacting dipole in a field of strength $E_f^\prime$, is $P_0(\zeta) \propto \exp(\beta E_f^\prime \zeta)$. Hence $E_f^\prime$ can be estimated by fitting $P_{0}$ to the numerically obtained distribution $P(\zeta)$. For all other values of $E$, we calculate the effective fields by linear scaling, $E^\prime=E E_f^\prime/E_f$. 
Again, using Eq. \ref{perfect} with $\bar H_{{ E}}({\bf z})= H_{{ E}^\prime,0}({\bf z})$ we obtain a new set of mapping functions. In particular, when the $E$ field is switched from $E_{i}$ to $E_{i+1}$, the $\zeta_k=\cos(\theta_k)$ degree of freedom of the $k^{th}$ dipole is transformed according to Eq. \ref{MeanFMdipole}
\begin{equation}
\label{MeanFMdipole}
m_i(\zeta_k)=\frac{1}{\beta E^\prime_{i+1}}{\ln\left [\frac{\sinh(\beta E^\prime_{i+1})}{\sinh(\beta E^\prime_i)} (e^{\beta E^\prime_i \zeta_k}-e^{\beta E^\prime_i} )+e^{\beta E^\prime_{i+1}}\right ]}
\end{equation}
We will refer to Eq.~\ref{MeanFMdipole} as a {\it mean field mapping}.
Since the single-dipole distributions for the interacting system at field strength $E$ are (by construction) closely approximated by the single-particle distributions for the non-interacting system at $E^\prime$, we expect the mean field mappings to perform better than the simple mappings of Eq. \ref{PerfectMdipole}.

We performed numerical simulations with $n_p=800$ particles. The parameters $\sigma$, $\epsilon$ of the Lennard-Jones potential set the length and the energy scale of the system, and we took $L=10\sigma$ and $T^*=k_B T/\epsilon=1$. Minimum image convention and periodic boundary conditions~\cite{Frenkel} were used. We performed $N_s = 10^4$ forward and reverse simulations to estimate the free energy difference between the ensembles corresponding to $E=0$ and $E=1$, switching the field strength in $N=10$ equal increments.
Ten Monte Carlo sweeps were performed between these updates in $E$. We obtained estimates of $\Delta F$ using:
(1) unescorted switching simulations (Eq.~\ref{NEW}),
(2) escorted simulations with the simple mappings (Eq.~\ref{PerfectMdipole}), and
(3) escorted simulations with the mean field mappings (Eq.~\ref{MeanFMdipole}).
For the latter, the effective fields were obtained as described in the previous paragraph. In particular, we found $E_f^\prime \approx 1.5 E_f$ and therefore we took $E_i^\prime = 1.5 E_i$ in Eq.~\ref{MeanFMdipole}.

Fig.~\ref{data2.his} shows the work distributions $P_F(W)$ and $P_R(-W)$ for these sets of simulations, and reveals a progression from virtually no overlap for the unescorted simulations, to some overlap for the simulations with the simple mappings, to nearly perfect overlap when using the mean field mappings.
This trend is in agreement with the expectations mentioned above, and provides direct evidence that the mappings we have constructed substantially reduce the lag and dissipation.
The first three rows of Table~\ref{Tab:2} quantify these observations.
In particular, row 3 gives the distance between the means of $P_F(W)$ and $P_R(-W)$, and shows that this hysteresis proceeds from nearly $250 k_BT$ to about $24 k_BT$ to less than $1 k_BT$ in the three cases.
Rows 4 to 6 illustrate the effect of this trend on the efficiency and accuracy of the free energy estimates.
The estimates of $\Delta F$ (that is, $\Delta F_F^{est}$, $-\Delta F_R^{est}$, and $\Delta F_{BAR}^{est}$) obtained from the unescorted simulations differ substantially from one another, indicating a high degree of bias.
The estimates corresponding to the simple mappings are markedly better, though they still suggest a degree of bias on the order of $1 k_BT$.
Finally, the simulations with the mean field mappings are in agreement to within about $0.05 k_BT$, indicating excellent accuracy and efficiency.
These findings are also in agreement with the values of the overlap integral $C$, shown in row 7.
This was too low to be estimated using the unescorted simulations, and approaches its maximal value of 1/2 when using the mean field mappings.
Using escorted simulations with the mean field mappings, with the acceptance ratio method (BAR), we found that we were able to generate estimates of $\Delta F$ with error bars on the order of $0.2 k_BT$, with about $N_s \sim 1/C^2 \sim 10 $ (data not shown).

\begin{table}
\begin{tabular}{|c|c|c|c|}
\hline
&No mapping&Mapping&Mean field mapping\\
\hline
$\langle W\rangle_F$&$-60.409 \pm 0.126$&$-177.074\pm 0.039$&$-189.079 \pm 0.010$\\
\hline
$\langle W\rangle_R$&$302.958 \pm 0.132$ &$200.607\pm 0.045$&$189.971 \pm 0.010$\\
\hline
$\langle W \rangle_F+\langle W\rangle_R$&$242.549 \pm 0.182$ &$23.533\pm 0.060$&$0.892 \pm 0.014$\\

\hline
$\Delta F^{est}_{F}$&$-114.189\pm 3.913$&$-187.612\pm 0.405$&$-189.552 \pm 0.011$\\
\hline
$\Delta F^{est}_{R}$&$262.232\pm 0.711$&$191.877\pm 0.310$&$189.502\pm 0.0140$\\
\hline
$\Delta F^{est}_{BAR}$&$-128.215\pm 3.324$&$-189.599\pm 0.110$&$-189.530\pm 0.008$\\
\hline
C& $\sim 0$ & $0.011 \pm 0.001$ &$0.407 \pm 0.001$\\
\hline
\end{tabular}
\caption{Estimates and Figures of Merit for $\gamma=0.1$. Note that the simulations with the mapping are much more efficient than those without. The forward and reverse work histograms obtained from the simulations without any mappings were so far apart that a reliable estimate of $C$ could not be obtained.}  
\label{Tab:2}
\end{table} 

\section{Summary}
\label{Conclusion}
Nonequilibrium fast switching estimates of free energy differences often perform poorly due to dissipation (see Fig \ref{lag.sketch}). The strategy developed here seeks to address this issue. By modifying the dynamics with additional terms that serve to escort the system along a near equilibrium trajectory and consequently reduce dissipation, we obtain efficient fast switching estimators (Eq. \ref{Map.discreteF}) for the free energy difference. The success of the strategy depends crucially on the choice of the mapping functions $M_i$:
the more effectively these reduce the dissipation, the more efficient the resulting estimator of $\Delta F$.

The examples presented in Section~\ref{Examples} illustrate this point. 
For the hard sphere solute, we used a simple mapping function that uniformly compresses the solvent, vacating the region into which the hard sphere expands (Eq.~\ref{Mapping}).
With this escorting function we were able to estimate $\Delta F$ directly from single-stage switching simulations, which would not have been feasible without escorting.
In the example of the Lennard-Jones dipole fluid, we used a reference system of non-interacting dipoles to construct a reasonable set of mapping functions (Eq.\ref{PerfectMdipole}), and then we further refined these mappings using mean field arguments (Eq.~\ref{MeanFMdipole}).
Figure~\ref{data2.his} and Table~\ref{Tab:2} illustrate the correlation between reduced dissipation and increased computational efficiency.
Because mean field theory often provides an good description of many-body systems, we speculate that this approach will prove effective for more complex problems of physical interest.

We have also discussed figures of merit, specifically the dissipation in the forward and reverse processes, and the overlap integral $C$ (Eqs.~\ref{NUnidirectional}, \ref{Relativeentropy}, \ref{Cdefine}).
For the two examples in Section~\ref{Examples}, we found that these quantities indeed track the effectiveness of the mapping functions.
This suggests that these figures of merit might be useful to iteratively improve the performance of the mapping functions.

Finally, the efficiency of our method might further be improved by applying it in combination with other methods, such as biased or umbrella sampling algorithms (see e.g.\ Refs~\cite{Zuckerman,Wu05,Seansun,Lagless1}).

\acknowledgements
We gratefully acknowledge useful discussions with Jordan Horowitz and Andy Ballard, and financial support from the National Science Foundation under CHE-0841557 and the University of Maryland, College Park.

\appendix
\section{Appendix}
Here we derive a relation between $N_s$ and $C$ for the bidirectional estimator, Eq. \ref{Bennett:FT}. The Bennett estimator, Eq. \ref{Bennett:FT} can be rewritten as a ratio of two free energy perturbation identities~\cite{Kollman}  
\begin{equation}
\frac{\langle  {P_H(W)}/{P_F(W)} \rangle _{P_F(W)}}{\langle  {P_H(W)}/{P_R(-W)} \rangle _{P_R(-W)}}=1
\end{equation}
where $\langle\dots\rangle_{P_F(W)}$ denotes an average over $W$ values sampled from $P_F(W)$, $\langle\dots\rangle_{P_R(-W)}$ denotes an average over $W$ values sampled from $P_R(-W)$, $P_H(W) \equiv C^{-1} \frac{P_F(W)P_R(-W)}{P_F(W)+P_R(-W)}$ with $ C= \int dW \frac{ P_F(W)P_R(-W)}{P_F(W)+P_R(-W)}$ is the normalized harmonic mean distribution. As the averages in the numerator and the denominator are over different ensembles, let us separately consider the number of the realizations required for each to converge.

The dominant contributions to the average in the numerator come from work values that are typically sampled from the harmonic mean distribution $P_H$~\cite{CJ:Rareevents}. The probability that these dominant values are observed in the forward process can be given by  $P=\int_{Typical} dW P_F(W)=\int_{Typical} dW P_H P_F(W)/P_H$, where $\int_{Typical}$ denotes that the integration is performed over the range of $W$ values that are typically sampled from the harmonic mean distribution ($P_H(W)$).

 Following Ref~\cite{CJ:Rareevents}, we now write
\begin{equation}
\label{approxN} 
P \sim \int_{Typical} dW P_H e^{\ln \frac{P_F}{P_H}} \sim e^{\langle \ln\frac{P_F}{P_H}\rangle_H}\int_{Typical} dW P_H\sim e^{\langle \ln\frac{P_F}{P_H}\rangle_H}
\end{equation}
The number of realizations $N_s$ required for adequate sampling can be roughly given by $N_s \sim P^{-1} \sim \exp{D[P_H||P_F]}$, where we have used $-\langle \ln\frac{P_F}{P_H}\rangle_H=D[P_H||P_F]$. The relative entropy $D[P_H|| P_F]$ satisfies the following inequality
\begin{equation}
\label{UpperboundRE}
\begin{split}
D[P_H||P_F]&=\int \frac{1}{C} \frac{ P_R P_F}{P_R+P_F} \ln{ \frac{P_R}{C(P_F+P_R)}} \\
& \leq \ln {\int \frac{1}{4 C^2}\frac{4 P_R^2P_F}{{(P_R+P_F)}^2}}\\
&\leq \ln  {\frac{1}{ 4 C^2} \int P_R}\\
&=-2 \ln 2 C\\
\end{split}
\end{equation}
where we have used the Jensen's inequality~\cite{Cover2006} for concave functions together with the identity $4 P_F P_R \leq {(P_F+P_R)}^2$. 
Finally, using Eq. \ref{UpperboundRE}, the number of realizations required to obtain a reliable estimate of $\Delta F$ using Bennett's method is bounded by
\begin{equation} 
\label{Nbennett}
N_s \leq \frac{1}{C^2}
\end{equation}
We have not included the numerical factors in the above relation as it is already an approximate equation.


\begin{thebibliography}{10}%
\makeatletter
\providecommand \@ifxundefined [1]{%
 \ifx #1\undefined \expandafter \@firstoftwo
 \else \expandafter \@secondoftwo
\fi
}%
\providecommand \@ifnum [1]{%
 \ifnum #1\expandafter \@firstoftwo
 \else \expandafter \@secondoftwo
\fi
}%
\providecommand \enquote [1]{``#1''}%
\providecommand \bibnamefont  [1]{#1}%
\providecommand \bibfnamefont [1]{#1}%
\providecommand \citenamefont [1]{#1}%
\providecommand\href[0]{\@sanitize\@href}%
\providecommand\@href[1]{\endgroup\@@startlink{#1}\endgroup\@@href}%
\providecommand\@@href[1]{#1\@@endlink}%
\providecommand \@sanitize [0]{\begingroup\catcode`\&12\catcode`\#12\relax}%
\@ifxundefined \pdfoutput {\@firstoftwo}{%
 \@ifnum{\z@=\pdfoutput}{\@firstoftwo}{\@secondoftwo}%
}{%
 \providecommand\@@startlink[1]{\leavevmode}%
 \providecommand\@@endlink[0]{}%
}{%
 \providecommand\@@startlink[1]{%
  \leavevmode
  \pdfstartlink
   attr{/Border[0 0 1 ]/H/I/C[0 1 1]}%
   user{/Subtype/Link/A<</Type/Action/S/URI/URI(#1)>>}%
  \relax
 }%
 \providecommand\@@endlink[0]{\pdfendlink}%
}%
\providecommand \url  [0]{\begingroup\@sanitize \@url }%
\providecommand \@url [1]{\endgroup\@href {#1}{\urlprefix}}%
\providecommand \urlprefix [0]{URL }%
\providecommand \Eprint[0]{\href }%
\@ifxundefined \urlstyle {%
  \providecommand \doi [1]{doi:\discretionary{}{}{}#1}%
}{%
  \providecommand \doi [0]{doi:\discretionary{}{}{}\begingroup
  \urlstyle{rm}\Url }%
}%
\providecommand \doibase [0]{http://dx.doi.org/}%
\providecommand \Doi[1]{\href{\doibase#1}}%
\providecommand \bibAnnote [3]{%
  \BibitemShut{#1}%
  \begin{quotation}\noindent
    \textsc{Key:}\ #2\\\textsc{Annotation:}\ #3%
  \end{quotation}%
}%
\providecommand \bibAnnoteFile [2]{%
  \IfFileExists{#2}{\bibAnnote {#1} {#2} {\input{#2}}}{}%
}%
\providecommand \typeout [0]{\immediate \write \m@ne }%
\providecommand \selectlanguage [0]{\@gobble}%
\providecommand \bibinfo [0]{\@secondoftwo}%
\providecommand \bibfield [0]{\@secondoftwo}%
\providecommand \translation [1]{[#1]}%
\providecommand \BibitemOpen[0]{}%
\providecommand \bibitemStop [0]{}%
\providecommand \bibitemNoStop [0]{.\EOS\space}%
\providecommand \EOS [0]{\spacefactor3000\relax}%
\providecommand \BibitemShut [1]{\csname bibitem#1\endcsname}%
\bibitem{Frenkel}%
  \BibitemOpen
  \bibfield{author}{%
  \bibinfo {author} {\bibfnamefont{D.}~\bibnamefont{Frenkel}}\ and\ \bibinfo
  {author} {\bibfnamefont{B.}~\bibnamefont{Smit}},\ }%
  \emph{\bibinfo {title} {Understanding Molecular Simulation}},\ \bibinfo
  {edition} {2nd}\ ed.\ (\bibinfo {publisher} {Academic Press},\ \bibinfo
  {address} {San Diego},\ \bibinfo {year} {2002})%
  \bibAnnoteFile{NoStop}{Frenkel}%
\bibitem{Chipot2007}%
  \BibitemOpen
  \bibfield{author}{%
  \bibinfo {author} {\bibfnamefont{C.}~\bibnamefont{Chipot}}\ and\ \bibinfo
  {author} {\bibfnamefont{A.}~\bibnamefont{Pohorille}},\ }%
  \emph{\bibinfo {title} {Free Energy Calculations}}\ (\bibinfo {publisher}
  {Springer, Berlin},\ \bibinfo {year} {2007})%
  \bibAnnoteFile{NoStop}{Chipot2007}%
\bibitem{CJ:Equality}%
  \BibitemOpen
  \bibfield{author}{%
  \bibinfo {author} {\bibfnamefont{C.}~\bibnamefont{Jarzynski}},\ }%
  \bibfield{journal}{%
  \bibinfo {journal} {Phys. Rev. Lett.}\ }%
  \textbf{\bibinfo {volume} {78}},\ \bibinfo {pages} {2690} (\bibinfo {year}
  {1997})%
  \bibAnnoteFile{NoStop}{CJ:Equality}%
\bibitem{CJ:MasterEquation}%
  \BibitemOpen
  \bibfield{author}{%
  \bibinfo {author} {\bibfnamefont{C.}~\bibnamefont{Jarzynski}},\ }%
  \bibfield{journal}{%
  \bibinfo {journal} {Phys. Rev. E}\ }%
  \textbf{\bibinfo {volume} {56}},\ \bibinfo {pages} {5018} (\bibinfo {year}
  {1997})%
  \bibAnnoteFile{NoStop}{CJ:MasterEquation}%
\bibitem{Kofke}%
  \BibitemOpen
  \bibfield{author}{%
  \bibinfo {author} {\bibfnamefont{D.~A.}\ \bibnamefont{Kofke}},\ }%
  \bibfield{journal}{%
  \bibinfo {journal} {Mol. Phys.}\ }%
  \textbf{\bibinfo {volume} {104}},\ \bibinfo {pages} {3701} (\bibinfo {year}
  {2006}),\ \bibinfo {note} {and references therein}%
  \bibAnnoteFile{NoStop}{Kofke}%
\bibitem{CJ:Rareevents}%
  \BibitemOpen
  \bibfield{author}{%
  \bibinfo {author} {\bibfnamefont{C.}~\bibnamefont{Jarzynski}},\ }%
  \bibfield{journal}{%
  \bibinfo {journal} {Phys. Rev. E}\ }%
  \textbf{\bibinfo {volume} {73}},\ \bibinfo {pages} {046105} (\bibinfo {year}
  {2006})%
  \bibAnnoteFile{NoStop}{CJ:Rareevents}%
\bibitem{Gore}%
  \BibitemOpen
  \bibfield{author}{%
  \bibinfo {author} {\bibfnamefont{J.}~\bibnamefont{Gore}}, \bibinfo {author}
  {\bibfnamefont{F.}~\bibnamefont{Ritort}},\ and\ \bibinfo {author}
  {\bibfnamefont{C.}~\bibnamefont{Bustamante}},\ }%
  \bibfield{journal}{%
  \bibinfo {journal} {Proc. Natl. Acad. Sci. U.S.A}\ }%
  \textbf{\bibinfo {volume} {100}},\ \bibinfo {pages} {12564} (\bibinfo {year}
  {2003})%
  \bibAnnoteFile{NoStop}{Gore}%
\bibitem{Lag1}%
  \BibitemOpen
  \bibfield{author}{%
  \bibinfo {author} {\bibfnamefont{D.~A.}\ \bibnamefont{Pearlman}}\ and\
  \bibinfo {author} {\bibfnamefont{P.}~\bibnamefont{Kollman}},\ }%
  \bibfield{journal}{%
  \bibinfo {journal} {J. Chem. Phys}\ }%
  \textbf{\bibinfo {volume} {91}},\ \bibinfo {pages} {7831} (\bibinfo {year}
  {1989})%
  \bibAnnoteFile{NoStop}{Lag1}%
\bibitem{Lag2}%
  \BibitemOpen
  \bibfield{author}{%
  \bibinfo {author} {\bibfnamefont{R.}~\bibnamefont{Wood}},\ }%
  \bibfield{journal}{%
  \bibinfo {journal} {J. Phys. Chem}\ }%
  \textbf{\bibinfo {volume} {95}},\ \bibinfo {pages} {4838} (\bibinfo {year}
  {1991})%
  \bibAnnoteFile{NoStop}{Lag2}%
\bibitem{Hermans91}%
  \BibitemOpen
  \bibfield{author}{%
  \bibinfo {author} {\bibfnamefont{J.}~\bibnamefont{Hermans}},\ }%
  \bibfield{journal}{%
  \bibinfo {journal} {J. Phys. Chem.}\ }%
  \textbf{\bibinfo {volume} {95}},\ \bibinfo {pages} {9029} (\bibinfo {year}
  {1991})%
  \bibAnnoteFile{NoStop}{Hermans91}%
\bibitem{SVCJ_lag}%
  \BibitemOpen
  \bibfield{author}{%
  \bibinfo {author} {\bibfnamefont{S.}~\bibnamefont{Vaikuntanathan}}\ and\
  \bibinfo {author} {\bibfnamefont{C.}~\bibnamefont{Jarzynski}},\ }%
  \bibfield{journal}{%
  \bibinfo {journal} {EPL (Europhysics Letters)}\ }%
  \textbf{\bibinfo {volume} {87}},\ \bibinfo {pages} {60005 (6pp)} (\bibinfo
  {year} {2009})%
  \bibAnnoteFile{NoStop}{SVCJ_lag}%
\bibitem{escorted}%
  \BibitemOpen
  \bibfield{author}{%
  \bibinfo {author} {\bibfnamefont{S.}~\bibnamefont{Vaikuntanathan}}\ and\
  \bibinfo {author} {\bibfnamefont{C.}~\bibnamefont{Jarzysnki}},\ }%
  \bibfield{journal}{%
  \bibinfo {journal} {Phys. Rev. Lett.}\ }%
  \textbf{\bibinfo {volume} {100}},\ \bibinfo {pages} {190601} (\bibinfo {year}
  {2008})%
  \bibAnnoteFile{NoStop}{escorted}%
\bibitem{CJ:Targeting}%
  \BibitemOpen
  \bibfield{author}{%
  \bibinfo {author} {\bibfnamefont{C.}~\bibnamefont{Jarzynski}},\ }%
  \bibfield{journal}{%
  \bibinfo {journal} {Phys. Rev. E}\ }%
  \textbf{\bibinfo {volume} {65}},\ \bibinfo {pages} {046122} (\bibinfo {year}
  {2002})%
  \bibAnnoteFile{NoStop}{CJ:Targeting}%
\bibitem{BijectiveTFEP}%
  \BibitemOpen
  \bibfield{author}{%
  \bibinfo {author} {\bibfnamefont{A.~M.}\ \bibnamefont{Hahn}}\ and\ \bibinfo
  {author} {\bibfnamefont{H.}~\bibnamefont{Then}},\ }%
  \bibfield{journal}{%
  \bibinfo {journal} {Phys. Rev. E}\ }%
  \textbf{\bibinfo {volume} {79}},\ \bibinfo {pages} {011113} (\bibinfo {year}
  {2009})%
  \bibAnnoteFile{NoStop}{BijectiveTFEP}%
\bibitem{miller:7035}%
  \BibitemOpen
  \bibfield{author}{%
  \bibinfo {author} {\bibfnamefont{M.~A.}\ \bibnamefont{Miller}}\ and\ \bibinfo
  {author} {\bibfnamefont{W.~P.}\ \bibnamefont{Reinhardt}},\ }%
  \bibfield{journal}{%
  \Doi{10.1063/1.1313537}{\bibinfo {journal} {The Journal of Chemical
  Physics}}\ }%
  \textbf{\bibinfo {volume} {113}},\ \bibinfo {pages} {7035} (\bibinfo {year}
  {2000})%
  \bibAnnoteFile{NoStop}{miller:7035}%
\bibitem{Crooksfluctuation}%
  \BibitemOpen
  \bibfield{author}{%
  \bibinfo {author} {\bibfnamefont{G.}~\bibnamefont{Crooks}},\ }%
  \bibfield{journal}{%
  \bibinfo {journal} {Phys. Rev. E}\ }%
  \textbf{\bibinfo {volume} {61}},\ \bibinfo {pages} {2361} (\bibinfo {year}
  {2000})%
  \bibAnnoteFile{NoStop}{Crooksfluctuation}%
\bibitem{Crooks1998}%
  \BibitemOpen
  \bibfield{author}{%
  \bibinfo {author} {\bibfnamefont{G.~E.}\ \bibnamefont{Crooks}},\ }%
  \bibfield{journal}{%
  \bibinfo {journal} {J. Stat. Phys.}\ }%
  \textbf{\bibinfo {volume} {90}},\ \bibinfo {pages} {1481 } (\bibinfo {year}
  {1998})%
  \bibAnnoteFile{NoStop}{Crooks1998}%
\bibitem{Crooks99}%
  \BibitemOpen
  \bibfield{author}{%
  \bibinfo {author} {\bibfnamefont{G.~E.}\ \bibnamefont{Crooks}},\ }%
  \bibfield{journal}{%
  \bibinfo {journal} {Phys. Rev. E}\ }%
  \textbf{\bibinfo {volume} {60}},\ \bibinfo {pages} {2721 } (\bibinfo {year}
  {1999})%
  \bibAnnoteFile{NoStop}{Crooks99}%
\bibitem{Bennett}%
  \BibitemOpen
  \bibfield{author}{%
  \bibinfo {author} {\bibfnamefont{C.~H.}\ \bibnamefont{Bennett}},\ }%
  \bibfield{journal}{%
  \bibinfo {journal} {J. Comput. Phys}\ }%
  \textbf{\bibinfo {volume} {22}},\ \bibinfo {pages} {245} (\bibinfo {year}
  {1976})%
  \bibAnnoteFile{NoStop}{Bennett}%
\bibitem{Bennett:Likelihood}%
  \BibitemOpen
  \bibfield{author}{%
  \bibinfo {author} {\bibfnamefont{M.~R.}\ \bibnamefont{Shirts}}, \bibinfo
  {author} {\bibfnamefont{E.}~\bibnamefont{Bair}}, \bibinfo {author}
  {\bibfnamefont{G.}~\bibnamefont{Hooker}},\ and\ \bibinfo {author}
  {\bibfnamefont{V.~S.}\ \bibnamefont{Pande}},\ }%
  \bibfield{journal}{%
  \bibinfo {journal} {Phys. Rev. Lett.}\ }%
  \textbf{\bibinfo {volume} {91}},\ \bibinfo {pages} {140601} (\bibinfo {year}
  {2003})%
  \bibAnnoteFile{NoStop}{Bennett:Likelihood}%
\bibitem{Note1}%
  \BibitemOpen
  \bibinfo {note} {As is usually the case with Monte Carlo simulations, we do
  not include momenta in the microstate.}%
  \bibAnnoteFile{Stop}{Note1}%
\bibitem{Kampen}%
  \BibitemOpen
  \bibfield{author}{%
  \bibinfo {author} {\bibfnamefont{N.~G.~V.}\ \bibnamefont{Kampen}},\ }%
  \emph{\bibinfo {title} {Stochastic Processes in Physics and Chemistry}}\
  (\bibinfo {publisher} {Elsevier},\ \bibinfo {address} {New York},\ \bibinfo
  {year} {2007})%
  \bibAnnoteFile{NoStop}{Kampen}%
\bibitem{Reinhardt1992}%
  \BibitemOpen
  \bibfield{author}{%
  \bibinfo {author} {\bibfnamefont{W.~P.}\ \bibnamefont{Reinhardt}}\ and\
  \bibinfo {author} {\bibfnamefont{J.~E.~Hunter.}\ \bibnamefont{III}},\ }%
  \bibfield{journal}{%
  \bibinfo {journal} {J. Chem. Phys}\ }%
  \textbf{\bibinfo {volume} {97}},\ \bibinfo {pages} {1599 } (\bibinfo {year}
  {1992})%
  \bibAnnoteFile{NoStop}{Reinhardt1992}%
\bibitem{Hunter1993}%
  \BibitemOpen
  \bibfield{author}{%
  \bibinfo {author} {\bibfnamefont{J.~E.~Hunter.}\ \bibnamefont{III}}, \bibinfo
  {author} {\bibfnamefont{W.~P.}\ \bibnamefont{Reinhardt}},\ and\ \bibinfo
  {author} {\bibfnamefont{T.~F.}\ \bibnamefont{Davis}},\ }%
  \bibfield{journal}{%
  \bibinfo {journal} {J. Chem. Phys}\ }%
  \textbf{\bibinfo {volume} {99}},\ \bibinfo {pages} {6856 } (\bibinfo {year}
  {1993})%
  \bibAnnoteFile{NoStop}{Hunter1993}%
\bibitem{Zuckerman02}%
  \BibitemOpen
  \bibfield{author}{%
  \bibinfo {author} {\bibfnamefont{D.~M.}\ \bibnamefont{Zuckerman}}\ and\
  \bibinfo {author} {\bibfnamefont{T.~B.}\ \bibnamefont{Woolf}},\ }%
  \bibfield{journal}{%
  \bibinfo {journal} {Phys. Rev. Lett.}\ }%
  \textbf{\bibinfo {volume} {89}},\ \bibinfo {pages} {180602} (\bibinfo {year}
  {2002})%
  \bibAnnoteFile{NoStop}{Zuckerman02}%
\bibitem{PhysRevLett.101.090602}%
  \BibitemOpen
  \bibfield{author}{%
  \bibinfo {author} {\bibfnamefont{E.~H.}\ \bibnamefont{Feng}}\ and\ \bibinfo
  {author} {\bibfnamefont{G.~E.}\ \bibnamefont{Crooks}},\ }%
  \bibfield{journal}{%
  \Doi{10.1103/PhysRevLett.101.090602}{\bibinfo {journal} {Phys. Rev. Lett.}}\
  }%
  \textbf{\bibinfo {volume} {101}},\ \bibinfo {pages} {090602} (\bibinfo
  {month} {Aug}\ \bibinfo {year} {2008})%
  \bibAnnoteFile{NoStop}{PhysRevLett.101.090602}%
\bibitem{Cover2006}%
  \BibitemOpen
  \bibfield{author}{%
  \bibinfo {author} {\bibfnamefont{T.~M.}\ \bibnamefont{Cover.}}\ and\ \bibinfo
  {author} {\bibfnamefont{J.~A.}\ \bibnamefont{Thomas}},\ }%
  \emph{\bibinfo {title} {Elements of Information Theory}}\ (\bibinfo
  {publisher} {Wiley-Interscience},\ \bibinfo {year} {2006})%
  \bibAnnoteFile{NoStop}{Cover2006}%
\bibitem{HahnThen2010}%
  \BibitemOpen
  \bibfield{author}{%
  \bibinfo {author} {\bibfnamefont{A.~M.}\ \bibnamefont{Hahn}}\ and\ \bibinfo
  {author} {\bibfnamefont{H.}~\bibnamefont{Then}},\ }%
  \bibfield{journal}{%
  \Doi{10.1103/PhysRevE.81.041117}{\bibinfo {journal} {Phys. Rev. E}}\ }%
  \textbf{\bibinfo {volume} {81}},\ \bibinfo {pages} {041117} (\bibinfo {month}
  {Apr}\ \bibinfo {year} {2010})%
  \bibAnnoteFile{NoStop}{HahnThen2010}%
\bibitem{Note2}%
  \BibitemOpen
  \bibinfo {note} {For $C<<1$, the upper and lower bounds in Eq. \ref {Cbound}
  can be orders of magnitude apart. Nevertheless, Eq. \ref {Cbound} can serve
  as a good consistency check for the quality of the estimates. For example an
  estimate of $\Delta F$ using Bennett's method from a data set of size
  $N_s\sim 10^6$ is reliable if $C\sim 0.001$.}%
  \bibAnnoteFile{Stop}{Note2}%
\bibitem{Note3}%
  \BibitemOpen
  \bibinfo {note} {An even better mapping would uniformly compress the entire
  region $r>R_i$, including the eight corners, onto the region $r>R_{i+1}$.
  However, due to the geometric mismatch between the spherical inner surface
  and cubic outer surface of these regions, such a mapping is not represented
  by a simple formula such as Eq.~\ref {Mapping}, and would need to be
  constructed numerically.}%
  \bibAnnoteFile{Stop}{Note3}%
\bibitem{Bootstrap}%
  \BibitemOpen
  \bibfield{author}{%
  \bibinfo {author} {\bibfnamefont{B.}~\bibnamefont{Efron}},\ }%
  \emph{\bibinfo {title} {The Jackknife, the Bootstrap and Other Resampling
  Plans}}\ (\bibinfo {publisher} {Society for Applied Mathematics},\ \bibinfo
  {year} {1982})%
  \bibAnnoteFile{NoStop}{Bootstrap}%
\bibitem{GoodPractices2010}%
  \BibitemOpen
  \bibfield{author}{%
  \bibinfo {author} {\bibfnamefont{A.}~\bibnamefont{Pohorille}}, \bibinfo
  {author} {\bibfnamefont{C.}~\bibnamefont{Jarzynski}},\ and\ \bibinfo {author}
  {\bibfnamefont{C.}~\bibnamefont{Chipot}},\ }%
  \bibfield{journal}{%
  \bibinfo {journal} {The Journal of Physical Chemistry B}\ }%
  \textbf{\bibinfo {volume} {114}},\ \bibinfo {pages} {10235} (\bibinfo {year}
  {2010})%
  \bibAnnoteFile{NoStop}{GoodPractices2010}%
\bibitem{Note4}%
  \BibitemOpen
  \bibinfo {note} {Of course, even after dividing the problem into stages, one
  can apply escorting by separately treating each stage as a switching
  simulation with one step, $N=1$, and using the mappings given by Eq.~\ref
  {Mapping}. We found that this further reduces the error bars by nearly a
  factor of six.}%
  \bibAnnoteFile{Stop}{Note4}%
\bibitem{Chandler}%
  \BibitemOpen
  \bibfield{author}{%
  \bibinfo {author} {\bibfnamefont{D.}~\bibnamefont{Chandler}},\ }%
  \emph{\bibinfo {title} {Introduction to Modern Statistical Mechanics}}\
  (\bibinfo {publisher} {Oxford University Press},\ \bibinfo {address} {New
  York},\ \bibinfo {year} {1987})%
  \bibAnnoteFile{NoStop}{Chandler}%
\bibitem{Zuckerman}%
  \BibitemOpen
  \bibfield{author}{%
  \bibinfo {author} {\bibfnamefont{F.~M.}\ \bibnamefont{Ytreberg}}\ and\
  \bibinfo {author} {\bibfnamefont{D.~M.}\ \bibnamefont{Zuckerman}},\ }%
  \bibfield{journal}{%
  \bibinfo {journal} {J. Chem. Phys}\ }%
  \textbf{\bibinfo {volume} {120}},\ \bibinfo {pages} {10876} (\bibinfo {year}
  {2004})%
  \bibAnnoteFile{NoStop}{Zuckerman}%
\bibitem{Wu05}%
  \BibitemOpen
  \bibfield{author}{%
  \bibinfo {author} {\bibfnamefont{D.}~\bibnamefont{Wu}}\ and\ \bibinfo
  {author} {\bibfnamefont{D.~A.}\ \bibnamefont{Kofke}},\ }%
  \bibfield{journal}{%
  \bibinfo {journal} {J. Chem. Phys.}\ }%
  \textbf{\bibinfo {volume} {122}},\ \bibinfo {pages} {204104} (\bibinfo {year}
  {2005})%
  \bibAnnoteFile{NoStop}{Wu05}%
\bibitem{Seansun}%
  \BibitemOpen
  \bibfield{author}{%
  \bibinfo {author} {\bibfnamefont{S.~X.}\ \bibnamefont{Sun}},\ }%
  \bibfield{journal}{%
  \bibinfo {journal} {J. Chem. Phys}\ }%
  \textbf{\bibinfo {volume} {118}},\ \bibinfo {pages} {5759} (\bibinfo {year}
  {2003})%
  \bibAnnoteFile{NoStop}{Seansun}%
\bibitem{Lagless1}%
  \BibitemOpen
  \bibfield{author}{%
  \bibinfo {author} {\bibfnamefont{D.~D.~L.}\ \bibnamefont{Minh}},\ }%
  \bibfield{journal}{%
  \bibinfo {journal} {jcp}\ }%
  \textbf{\bibinfo {volume} {130}} (\bibinfo {year} {2009})%
  \bibAnnoteFile{NoStop}{Lagless1}%
\bibitem{Kollman}%
  \BibitemOpen
  \bibfield{author}{%
  \bibinfo {author} {\bibfnamefont{R.~J.}\ \bibnamefont{Radmer}}\ and\ \bibinfo
  {author} {\bibfnamefont{P.}~\bibnamefont{Kollman}},\ }%
  \bibfield{journal}{%
  \bibinfo {journal} {J. Comp. Chem}\ }%
  \textbf{\bibinfo {volume} {18}},\ \bibinfo {pages} {902} (\bibinfo {year}
  {1997})%
  \bibAnnoteFile{NoStop}{Kollman}%
\end{thebibliography}
\end{document}